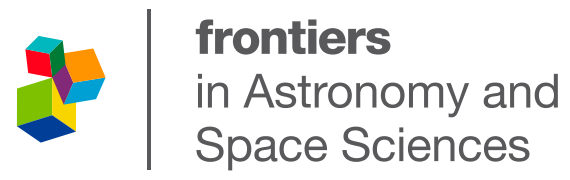

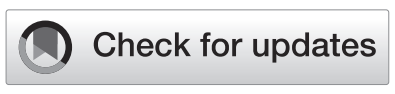

# Gravitational instability in planet-forming discs

Cristiano Longarini[1]* and Giuseppe Lodato[2]

[1]Institute of Astronomy, University of Cambridge, Cambridge, United Kingdom, [2]Dipartimento di Fisica Aldo Pontremoli, Università degli Studi di Milano, Milano, Italy

*CORRESPONDENCE
Cristiano Longarini,
cl2000@cam.ac.uk

During the earliest phases of star and planet formation, a young star is surrounded by a disc that can become gravitationally unstable. Indeed, if the protoplanetary disc is sufficiently massive, its self-gravity triggers the formation of large scale spiral arms, transporting angular momentum, trapping solid particles and, potentially, collapsing to form sub-stellar companions. In this paper, we review the theoretical and observational advances in the study of gravitational instability in protoplanetary discs over the last 10 years, since the advent of ALMA. These developments have transformed our understanding of gravitational instability, moving beyond its historical role as a theoretical mechanism for giant planet formation toward a physically rich framework that can be directly tested against high-resolution observations.

## 1 Introduction

Protoplanetary discs are the site of planet formation, providing the physical link between protostars and the emergence of planetary systems. Constraining their dynamical properties is therefore essential to understand how planets form, how efficiently material is converted into planets, and what types of planetary systems are expected to emerge.

During the last decade, the observational landscape has been revolutionised by the Atacama Large Millimetre Array (ALMA), which has enabled spatially and spectrally resolved observations of gas and dust in protostellar environments (Andrews et al., 2018), uncovering their morphology and the physical processes taking place within them. In particular, high-resolution continuum and molecular line observations have revealed a rich variety of substructures in protoplanetary discs (Andrews, 2020), including rings, gaps (ALMA Partnership et al., 2015; Huang et al., 2018a; Clarke et al., 2018; Guzmán et al., 2018; Long et al., 2018), spirals (Pérez et al., 2016; Huang et al., 2018b), and asymmetries (Casassus et al., 2013; Isella et al., 2013; Marino et al., 2015; Curone et al., 2025), opening new windows onto the dynamics of planet-forming systems.

In this paper, we focus on one of the most fundamental aspects of disc physics: the role of the disc self-gravity. Protoplanetary disc mass sets the total reservoir available for planet formation, regulates the aerodynamical coupling between gas and dust, and determines whether key physical processes, such as hydrodynamical instabilities, can operate. In particular, when the disc becomes sufficiently massive, its self-gravity (SG) can no longer be neglected. In this regime, the gravitational potential of the disc becomes comparable to that of the central star, fundamentally modifying its dynamics. One of the most important consequences is the onset of gravitational instability (GI), a mechanism that can drive angular momentum transport, generate large-scale spiral structures, and potentially lead to fragmentation under the right physical conditions.

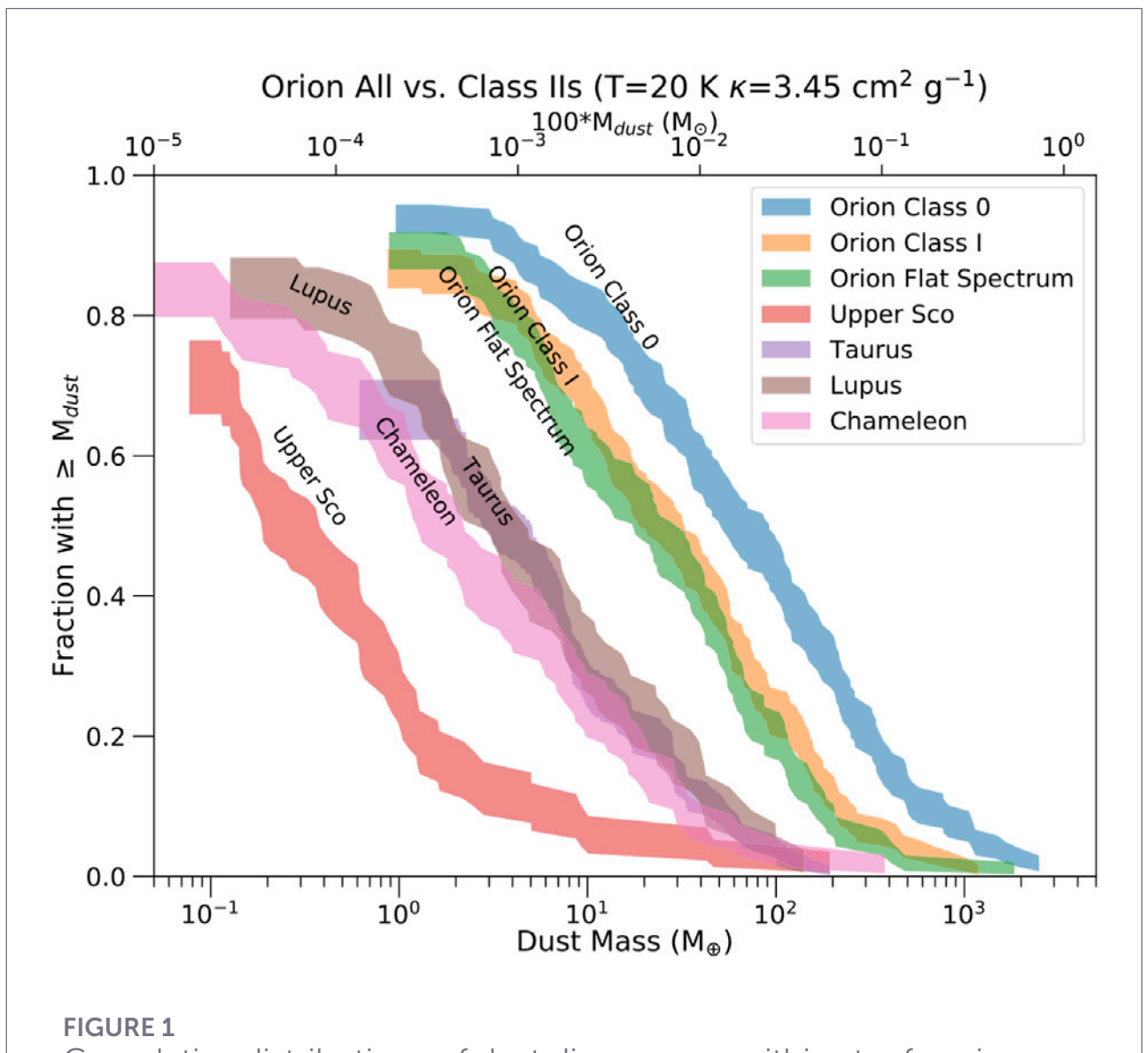

FIGURE 1
Cumulative distributions of dust disc masses within star forming regions at different ages (taken from Tobin et al., 2020). The plot clearly shows a modulation with age, where younger star forming regions (i.e., Orion) harbour more massive discs compared to more evolved ones (i.e., Upper Sco). This suggests that the role of the disc self-gravity should be considered in the early stages of planet formation.

Gravitational instability is expected to be especially relevant in the earliest stages of disc evolution. Young discs are likely more massive, as protoplanetary discs are accretion discs whose mass decreases over time due to disc dispersal processes, such as viscous accretion or disc winds (Pringle, 1981; Blandford and Payne, 1982). Hydrodynamical simulations of star formation, that follow the collapse of molecular clouds (Bate, 2018; Lebreuilly et al., 2024) typically result in the formation of star + disc systems, where the disc mass is a substantial fraction of the total mass. This trend is supported observationally by comparisons of disc mass functions[1] across star-forming regions of different ages: younger regions host systematically more massive discs, while more evolved regions are dominated by lower-mass systems (Tobin et al., 2020), as shown in Figure 1. This makes SG and GI not only a theoretical possibility, but a key ingredient in understanding these early stages.

The very early stages of protoplanetary disc lifetime are fundamental to understand the process of planet formation. Direct imaging of protostellar discs, in scattered light and especially at sub-mm wavelengths, has opened debate regarding the timescale for planet formation. Many of the large and bright discs observed with ALMA and SPHERE show substructures (Andrews et al., 2018; Andrews, 2020) that is consistent with, and often interpreted as, the theoretical signpost of planet-disc interaction (Dong et al., 2015b,a; Lodato et al., 2019; Paardekooper et al., 2023). A subset of these discs, including HL Tau (ALMA Partnership et al., 2015), IRS 63 (Segura-Cox et al., 2020) and GY91 (Sheehan and Eisner, 2018) are young, with ages estimated to be less than 1 Myr. Assuming a planet origin for the substructure, a substantial part of the planet formation process must overlap with the time when protostellar discs are still young, and likely self-gravitating.

In this paper, we focus on the physics of gravitational instability in young protostellar discs, combining theoretical developments and observational advances over the past decade. The paper is structured as follows. In Section 2, we provide a brief overview of gravitational instability in protostellar discs. In Section Equation 3, we present recent theoretical progress, focusing on numerical developments, the connection between GI and dust dynamics, planet formation, and thermodynamics. In Section 4, we discuss observational advances, highlighting three sources as representative examples of GI in discs, and review methods to infer disc masses in bright systems. Finally, in Section 5, we summarise our conclusions. Gravitational instability in protoplanetary discs has been the subject of extensive study, and we refer the reader to the comprehensive review by (Kratter and Lodato, 2016) for a thorough overview. The goal of this work is to present the developments and advances that have emerged since 2016.

## 2 Summary of self gravity and gravitational instability in protostellar discs

Self-gravity is fundamental in astrophysics, from cosmic structures to stars and planetary rings. In the context of protoplanetary discs, self-gravity is particularly important in the early stages of the disc lifetime, when the disc-to-star mass ratio is high. A crucial effect of the disc self-gravity is the development of gravitational instability. This mechanism is characterised by the formation of large scale spiral structures, transporting angular momentum throughout the disc. The foundations of our understanding of self-gravity and gravitational instability in discs comes from the field of galactic dynamics. Indeed, during the second half of the last century, the so-called "density wave theory" was developed to explain the origin of spiral structure in galaxies (Lin and Shu, 1964; Toomre, 1964). In this context, massive protoplanetary discs can be treated, to first approximation, in analogy with galactic discs, provided that the appropriate rescaling of physical quantities and gravitational potentials is taken into account.

In this section, we provide a concise overview of the main properties of gravitational instability in protostellar discs. We introduce the instability criterion and the key physical quantities that characterise the GI regime. This is not intended to be a comprehensive review of the subject, but rather a minimal framework to support the discussion in the following sections.

### 2.1 Linear theory of gravitational instability

To introduce the linear behaviour of gravitational instability, it is worth discussing the dispersion relation, i.e., how the perturbation frequency $\omega$ depends on the radial wavenumber $k$. This relationship was first obtained in the context of galactic dynamics by Safronov (1960), and then formalised by Lin and Shu (1964). In the razor thin

1 The trend presented in Tobin et al. (2020) is derived from dust mass measurements based on mm-flux observations. Measuring protoplanetary disc masses—especially in young systems—remains highly challenging; here, we assume that the trend in total disc masses mirrors that observed in dust masses.

disc limit, in the WKB approximation, the dispersion relation reads

$$(\omega - m\Omega)^2 = c_s^2 k^2 - 2\pi G\Sigma|k| + \kappa^2, \tag{1}$$

where $m$ is the number of spiral arms, $c_s$ the sound speed of the fluid, $\Sigma$ the disc surface density and $\kappa$ the epicyclic frequency, that for a Keplerian disc $\kappa = \Omega_k = \sqrt{GM_\star/R^3}$. The dispersion relation is a quadratic in $k$, and each term has a physical interpretation. The left side term is the Doppler shifted perturbation frequency, and can be re-written by underlying the spiral pattern frequency $\omega - m\Omega = m(\Omega_p - \Omega)$, where $\Omega_p = \omega/m$ is the spiral pattern frequency. The first term on the right side of Equation 1 is the stabilizing pressure contribution. It is quadratic in $k$ and hence it is significant for large wavenumbers - short wavelengths. The second term is the disc self-gravity, that is linear with the wavenumber. Its contribution is more significant at intermediate wavelengths, and its role is destabilizing, contributing with a negative sign. Finally, the third term is related to the disc rotation and represents epicyclic oscillations. It does not depend on $k$ and hence its stabilizing effect is dominant at large wavelengths. The competition of these three effects determines the stability of the system. This is particularly clear when studying the instability threshold for axisymmetric perturbations, namely, $\omega^2 = 0$. The perturbation is unstable if

$$Q = \frac{c_s\kappa}{\pi G\Sigma} < 1, \tag{2}$$

where $Q$ is the so-called Toomre parameter (Toomre, 1964). The instability threshold is hence $Q = 1$. In a protostellar disc, the sound speed and the surface density can be described as power laws with radius. We usually assume that $c_s \propto R^{-q/2}$, and $\Sigma \propto R^{-\gamma}$: this implies that $Q \propto R^{\gamma - q/2 - 3/2}$. If $\gamma < (q+3)/2$, the Toomre parameter decreases with the radius, meaning that the outer part of the disc is more prone to be unstable. Conversely, if $\gamma > (q+3)/2$, the disc will be more likely gravitationally unstable in the inner regions. If we consider a standard power-law exponent for temperature ($q = 1/2$), it is likely that the condition for instability will be met in the outer (inner) regions of the disc when $\gamma < 7/4$ ($\gamma > 7/4$). Observations of discs have typically yielded power-law exponents within the range of [0, 2] (Andrews and Williams, 2007), making it more likely for the condition $Q < 1$ to be met in the outer regions of a disc.

For the case of an unstable disc, the most unstable wavenumber $k_J$ (i.e., where the right-hand side of Equation 1 is at its minimum) occurs at

$$k_J = \frac{\pi G\Sigma}{c_s^2} = H_{SG}^{-1}. \tag{3}$$

When the disc is marginally unstable ($Q = 1$), only modes close to $k_{uns}$ are excited. If we evaluate Equation 1 for $Q = 1$ and $k = k_J$, we get the condition $\Omega_p = \Omega$, which means that all excited modes are expected to be close to co-rotation. It is worth noting that the most unstable wavenumber $k_J$ is exactly the inverse of the disc thickness in the self-gravitating limit $H_{SG}$. Equivalently, it is possible to define the most unstable wavelength $\lambda_J = 2\pi/k_J$. $k_J$ and $\lambda_J$ are respectively the most unstable wavenumber and wavelength, often called Jeans wavenumber and wavelength.

## 2.2 Non-linear evolution of gravitational instability

Now, we ask what is the non-linear evolution of GI, and whether it can lead to transport of angular momentum. To tackle this question, it becomes imperative to employ numerical simulations of gravitationally unstable discs. Over the past 2 decades, this topic has gained considerable attention, employing various numerical techniques, ranging from global 3D SPH simulations (Rice et al., 2003, 2004; Lodato and Clarke, 2011), global grid based simulations (Kratter and Matzner, 2006; Kratter et al., 2008; Meru and Bate, 2010; 2012), local grid simulations (Booth and Clarke, 2019; Leedham et al., 2025; Rice et al., 2025).

A useful starting point is the Toomre stability parameter, $Q$, which scales proportionally with the sound speed (and hence temperature) and inversely with the disc surface density. As a consequence, cold and massive discs are more prone to gravitational instability. This naturally identifies two primary pathways to trigger GI: a reduction in temperature (cooling-driven GI) or an increase in surface density (infall- or mass-loading-driven GI). A morphological comparison between these two regimes is shown in Figure 2.

The non-linear outcome of GI broadly falls into two regimes. In the first, the disc settles into a self-regulated state, in which spiral density waves reach a quasi-steady amplitude and transport angular momentum outward. In this regime, for relatively light discs, GI can be described in terms of an effective turbulent viscosity (Lodato and Rice, 2004), often parameterised through an $\alpha$ prescription. In the second regime, the instability grows exponentially, leading to fragmentation: spiral perturbations collapse into gravitationally bound clumps (Gammie, 2001; Rice et al., 2003), potentially giving rise to the formation of companions (Boss, 1997).

In the following, we first discuss the self-regulated regime, highlighting the differences between cooling-driven and infall-driven GI, and then examine the conditions under which fragmentation occurs.

### 2.2.1 Cooling-driven

Self-regulation can be achieved through cooling. Let us consider an initially hot and gravitationally stable disc ($Q \gg 1$) that progressively cools. As the temperature decreases, the value of $Q$ correspondingly drops, eventually approaching the marginal stability threshold ($Q \sim 1$). At this point, gravitational instability turns on: the disc develops spiral density structures which, through compression and shocks, dissipate energy and heat up the gas.

This establishes a feedback loop in which cooling drives the disc toward instability, while GI-induced heating counteracts further cooling (Bertin, 2000). As a result, the disc evolves toward a quasi-steady state in which heating and cooling balance each other, maintaining $Q$ close to unity. In this sense, gravitational instability acts as a thermostat, activating efficient angular momentum transport and dissipation only when the disc becomes sufficiently cold.

We can model the cooling through the $\beta$–parameter, that is the cooling timescale in unit of the dynamical one, defined as $\beta = \Omega t_{cool}$. Cossins et al. (2009) showed that in the self-regulated state, the amplitude of the spiral perturbation defined as $\delta\Sigma/\Sigma$ is solely determined by the $\beta$–cooling, according to

$$\frac{\delta\Sigma}{\Sigma} \simeq \beta^{-1/2}. \tag{4}$$

Faster cooling (i.e., lower values of the cooling parameter $\beta$) leads to larger-amplitude spiral perturbations. In order to maintain thermal balance, the disc must generate stronger gravitational stresses to compensate for the increased cooling rate. This results in more prominent spiral structures and enhanced angular momentum transport.

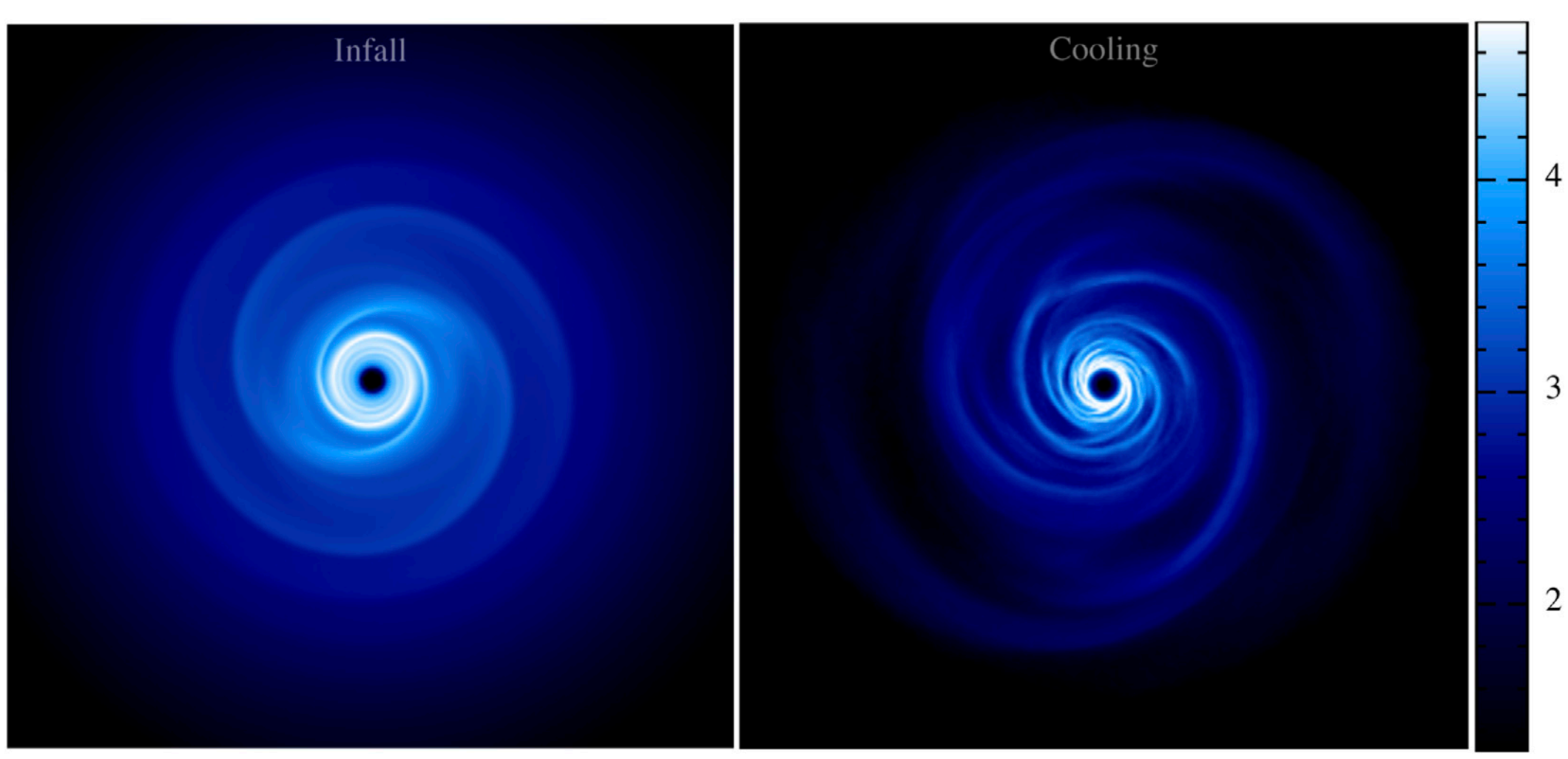

FIGURE 2
Hydrodynamical simulations of cooling driven and infall driven gravitational instability, from Longarini et al. (2025b). The two scenarios shows morphological analogies and differences. While the dominant mode in the two cases is the same ($m = 2$), the design of the spirals significantly differs, showing a clean, grand design spiral in the infall case, against a more flocculent spiral in the cooling one.

The self-regulated state described above can be quantified by relating the strength of angular momentum transport to the cooling rate. In particular, if no additional heating sources are present, thermal equilibrium requires that the heating provided by gravitational stresses balances radiative cooling. Under this assumption, the viscous heating timescale must match the cooling timescale, i.e., $t_{cool} = t_{th}$. This condition leads to a direct relation between the effective viscosity and the cooling time. The resulting $\alpha$ parameter associated with gravitational instability can be written as (Gammie, 2001; Lodato and Rice, 2004)

$$\alpha_{GI} = \frac{4}{9\gamma(\gamma - 1)\beta}. \quad (5)$$

This expression makes explicit that faster cooling (lower $\beta$) requires stronger gravitational stresses, and therefore a higher effective $\alpha$, to maintain thermal balance. This provides a quantitative framework for the self-regulated state described above.

### 2.2.2 Infall-driven

Self-regulation can also be achieved in discs that are continuously fed by mass infall from the surrounding envelope. In this case, instead of reducing the temperature, gravitational instability is triggered by an increase in the disc surface density. As mass is injected into the disc, the Toomre parameter $Q$ decreases, eventually approaching the marginal stability threshold ($Q \sim 1$), at which point GI develops.

Similarly to the cooling-driven case, the onset of GI leads to the formation of spiral density structures that transport angular momentum and redistribute mass within the disc. This transport acts to process the incoming material, preventing an indefinite build-up of mass and driving the system toward a quasi-steady state. In this sense, infall-driven GI can also establish a self-regulated configuration, in which the disc adjusts its structure to accommodate the imposed mass accretion rate.

A useful framework to characterise this regime was introduced by Kratter et al. (2010), who defined two dimensionless parameters describing the strength of accretion relative to the disc dynamical properties:

$$\xi = \frac{G\dot{M}_{inf}}{c_s^3}, \qquad \Gamma = \frac{\dot{M}_{inf}}{M_{\rm T}\Omega}, \quad (6)$$

where $\dot{M}_{inf}$ is the infall rate, $c_s$ is the sound speed, $M_{\rm T}$ is the total system mass (disc plus central object), and $\Omega$ is the local orbital frequency. The parameter $\xi$ measures the infall rate in units of the isothermal collapse rate, while $\Gamma$ compares the mass-loading timescale to the dynamical timescale of the system.

Using three-dimensional simulations of discs undergoing self-similar accretion, Kratter et al. (2010) showed that the system can either settle into a quasi-steady, self-regulated state or undergo fragmentation, depending on the values of these parameters.

More recently, the nature of self-regulation in infall-driven discs has been revisited in Longarini et al. (2025b), where it is shown that continuous mass loading can maintain gravitational instability over extended timescales, establishing a tight coupling between disc mass, stellar mass, and accretion rate. We defer a detailed discussion of this regime to Section 3.5.

### 2.2.3 GI as a companion formation scenario - gas fragmentation

If the disc cools too efficiently, or receives too much mass from the surrounding envelope, gravitational instability cannot reach a self-regulated state, resulting in the exponential growth of spiral perturbations. In this regime, the spiral arms fragment into gravitationally bound objects (Durisen et al., 2007).

In the cooling driven scenario, the critical value of $\beta$ below which fragmentation occurs has been the subject of extensive debate over the past 2 decades. Firstly, Gammie (2001) performed shearing box simulations of gravitational instability using the $\beta$−cooling prescription. He found that if the cooling timescale is faster than approximately 3 times than the dynamical one, fragmentation occurs. Afterwards, Rice et al. (2005) tackled this problem through three-dimensional global SPH simulations. They found that, depending on the value of the $\gamma$ coefficient, the critical value of $\beta$ is between 6 and 10. Meru and Bate (2010) and Meru and Bate (2011) performed high resolution SPH simulations, and they did not find convergence in their results when using a different number of gas particles. Lodato and Clarke (2011) proposed that this behaviour could be driven by resolution effects. One year later, Meru and Bate (2012) obtained evidence that convergence with increasing resolution occurs with SPH, as long as the effects of numerical viscosity are taken into account. In particular, the authors found that reducing the dissipation from the numerical viscosity leads to larger values of the critical cooling time at a given resolution. This happens because the numerical dissipation in SPH depends on resolution. Additionally, they also found convergence for grid based codes, however the results were in contrast with the particle based ones. From this analysis, they found that $\beta_{\rm crit}$ is larger than previously thought. In particular, for $\gamma = 5/3$, their critical $\beta$ was higher than 20, implying that young discs are very likely to fragment. Finally, Deng et al. (2017) obtained convergence between SPH and grid based codes, and the value of $\beta$ below which fragmentation occurs is

$$\beta_{\rm crit} = 3, \tag{7}$$

as firstly proposed by Gammie (2001). A minimum value of $\beta$ implies a maximum value of $\alpha$, i.e., the maximum amount of angular momentum that can be transported through GI. For $\beta_{\rm crit} = 3$, the maximum value of $\alpha$ is

$$\alpha_{\rm max} = 0.13, \tag{8}$$

for $\gamma = 5/3$.

In the infall-driven regime, Kratter et al. (2010) showed that the dimensionless parameters $\xi$ and $\Gamma$ are strong predictors of fragmentation. In particular, fragmentation is found to occur when

$$\Gamma \gtrsim \frac{\xi^{2.5}}{850}. \tag{9}$$

In the '90s, the discovery of the first extrasolar giant planet (Mayor and Queloz, 1995) presented a significant challenge to theorists, as the widely accepted core accretion model could not explain the formation of such a massive planet, because of the too long timescale. Boss (1997) proposed gravitational instability as an alternative paradigm for giant planet formation. For disc-to-star mass ratios of the order of $\gtrsim 0.1$ (i.e., comparable to the disc aspect ratio $H/R$), the outer part of the disc is likely to be marginally gravitationally unstable, according to the $Q$-criterion, for a power-law surface density profile. For a more realistic surface density profile, e.g., a self-similar one, a somewhat higher disc-to-star mass ratio, $M_d/M_* \gtrsim 0.25$, is typically required to reach $Q \sim 1$ in the outer disc, for the same aspect ratio. If the thermal saturation is not reached, because of a too short cooling time, gas fragmentation can happen. Boss (1997) proposed gas fragmentation as a way to rapidly form planets in the outer disc. One natural question to ask is: what is the typical mass of a GI fragment? We expect that the mass of the fragment is of the order of the Jeans mass. According to the quadratic dispersion relationship, the Jeans mass is

$$M_J = \Sigma\lambda_J^2 = 4\pi\left(\frac{H}{r}\right)^4 \frac{M_\star}{M_d} M_\star = 10^{-2}\left(\frac{H/r}{0.1}\right)^4 \left(\frac{M_d/M_\star}{0.1}\right)^{-1} M_\star; \tag{10}$$

this means that for a solar mass star, the typical Jeans mass is of the order of $\sim 10 M_{\rm jup}$. It is important to note that this value is just an initial mass. Indeed, the GI fragment will start accreting material in the disc, that is dense because the disc is massive. This process will lead to runaway accretion, and to the formation of a stellar companion (Kratter and Matzner, 2006) rather than a planet. Indeed, the theoretical minimum mass a star must have to undergo fusion at the core, is estimated to be about $75 M_{\rm jup}$, and it can be rapidly reached in those conditions. It is worth noting that the historical motivation for gravitational instability, namely, the difficulty of core accretion in forming massive cores within the disc lifetime, has been substantially reconsidered in light of more recent developments in the core accretion paradigm (e.g., Venturini et al., 2020; Drkazkowska et al., 2023. While this is true in the inner disc, the standard core accretion scenario still struggles in explaining planet formation in the outer discs, in cases such WISPIT2 (van Capelleveen et al., 2025) and HR 8799 (Marois et al., 2008). Although it circumvents the timescale problem, the standard gravitational instability scenario is not likely to explain the formation of planets, without invoking *ad hoc* mechanisms, such as tidal downsizing Nayakshin (2010).

# 3 Theoretical advances on GI

## 3.1 GI wiggle and kinematics

During the last 10 years, ALMA and SPHERE data showed a huge number of protoplanetary discs with spiral structures (Huang et al., 2018b). The origin of such structures remains a subject of debate. While planet-disc interactions, stellar flybys, and gravitational instability can all produce similar morphologies, gas kinematics offers a powerful diagnostic to break this degeneracy. Specifically, the kinematic signatures of GI are expected to be distinct and clearly identifiable.

Using hydrodynamics simulations coupled with radiative transfer, Hall et al. (2020) first demonstrated that a disc undergoing GI exhibits clear kinematic signatures in molecular line observations. This feature, referred to as the "GI Wiggle", manifests as a distinctive zig-zag pattern in the emission lines (isovelocity contours), most commonly seen in the CO roto-vibrational lines in the mm band. A key property that distinguishes the GI wiggle from planet-driven perturbations (e.g., Bollati et al., 2021) is its global nature. While a planet induces a localised perturbation that decays with distance from the source, GI-induced spirals perturb the disc across its entire radial and azimuthal extent, resulting in a coherent, large-scale kinematic signal.

Longarini et al. (2021) provided an analytical description of the GI wiggles in the linear regime of GI, under the assumption of self-regulation. As a wave-like perturbation, the wiggle is characterised by both its wavelength and its amplitude. Since the kinematic perturbation follows the density enhancement of the spiral arms, the number of arms and their pitch angle directly determine the spatial frequency and the position of the wiggles in the velocity field. These parameters are not arbitrary; they scale with the disc-to-star mass ratio, with more massive discs typically displaying fewer and more open spiral arms (i.e., larger pitch angles). The amplitude of the GI wiggle is a particularly critical quantity since it allows to constrain fundamental disc properties. Longarini et al. (2021) showed that this quantity is determined by the disc-to-star mass ratio and the spiral density contrast, that is related to the angular momentum transported by the spirals, according to

$$\mathcal{A}_v = 3m\sqrt{\alpha_{\rm GI}\gamma(\gamma-1)}(M_d/M_\star)^2 v_k, \tag{11}$$

where $\alpha_{GI}$ measures the amount of angular momentum transported by gravitational instability, and $v_k$ is the Keplerian azimuthal velocity.

In a self-regulated state, more massive discs are typically hotter[2], leading to a higher sound speed. This results in an increased perturbation amplitude, as the kinematic response is proportional to the local sound speed. Furthermore, the velocity perturbation scales with the normalised density fluctuation, $\delta\Sigma/\Sigma$. This quantity reflects the specific mechanism triggering the instability (as shown in Equation 4), but it is most fundamentally related to the gravitational stress exerted by the spirals. When the disc reaches a self-regulated state where angular momentum transport can be described via a pseudo-viscous $\alpha$-prescription, the following relation holds

$$\frac{\delta\Sigma}{\Sigma} \propto \sqrt{\alpha_{eff}}, \tag{12}$$

where $\alpha_{eff}$ is the effective $\alpha$ viscosity produced by gravitational instability. This connection is profound: it implies that by measuring the amplitude of the kinematic "wiggles" in molecular line observations, we can potentially "weigh" the disc and simultaneously measure the efficiency of its internal transport mechanisms.

However, the applicability of this analytical framework to real observations depends on whether the disc operates within the linear regime of the instability. While the linear theory provides a clear link between kinematics and disc properties, its validity must be carefully verified against global, non-linear models. Figure 3 illustrates this by comparing the analytical predictions of the wiggle with synthetic observations derived from 3D hydrodynamical simulations (Speedie et al., 2024). Although GI is an inherently non-linear process characterised by shocks and strong density gradients, the large-scale kinematic response retains the features predicted by the linear theory. A further test of this analytical framework is provided in Terry et al. (2022), where the expected scaling of the wiggle amplitude with disc mass is remarkably well recovered in numerical simulations. By measuring the velocity deviations in synthetic observations of discs with varying masses, they confirmed that the kinematic response remains sensitive to the underlying gravitational potential even as the instability enters the non-linear regime.

2 This is a direct consequence of the marginal stability condition $Q = 1$. Since $Q \sim (M_\star/M_d)(H/R)$, the self-regulation of the instability effectively sets the disc's thermal structure, leading to $H/R \sim M_d/M_\star$.

It is important to underline that identifying these kinematic signatures from observations, and disentangling them from other kinds of signatures, is not straightforward. Indeed, high spatial and spectral resolution ALMA observations are required, and the geometry of the disc—such as its extent and its inclination with respect to the line of sight—plays an important role. For these reasons, the number of sources for which this method is currently applicable remains very small.

Altogether, the GI wiggle stands out as a unique kinematic signature, encoding in a single observable the disc mass, the spiral morphology, and the efficiency of gravitational stress. This makes kinematics an indispensable tool in the interpretation of self-gravitating discs.

## 3.2 GI and dust, planetary cores formation

Planet formation through gravitational instability is often synonymous with gas fragmentation, however this process typically results in the formation of stellar-mass companions or brown dwarfs rather than planetary-mass objects (Kratter and Lodato, 2016). Nevertheless, the role of GI in planet formation extends far beyond the direct collapse of the gaseous component. In this section, we explore the interplay between dust and gravitational instability, highlighting how GI-induced structures can alter the evolution of the solid component and provide a rapid pathway for the formation of planetary cores.

Rice et al. (2004) first proposed and demonstrated that gas spiral arms act as efficient dust traps. As pressure maxima, these structures naturally collect dust particles, with the trapping process being particularly effective for grains with a Stokes number ($St = t_s\Omega$, where $t_s$ is the stopping time) close to unity. Contrary to planetary-driven spirals, GI spirals efficiently trap dust particles because their pattern speed is close to corotation (Cossins et al., 2009), making the relative velocity between the pattern and the dust particles very small, whereas planetary spirals propagate at a fixed pattern speed. Beyond simple concentration, the authors suggested that the dust density within these arms could reach a critical threshold, leading to the direct gravitational collapse of the solid component into planetary cores. This hypothesis was further explored in Rice et al. (2006), confirming the viability of the mechanism (left panel of Figure 4). While those early numerical simulations were limited by resolution and could not fully resolve the dust Jeans length, they opened a new avenue for planet formation in young, massive discs: the collapse of dust within spiral structures. This *hybrid* scenario combines the advantages of both traditional theories, offering a mechanism that is as fast and efficient as gravitational instability, while being fundamentally dust-based like core accretion.

To understand the physics behind this process, it is necessary to consider the gravitational stability of a two-fluid disc. This problem has been thoroughly explored in the context of galactic dynamics, where the two components are typically stars and gas (Jog and Solomon, 1984; Bertin and Romeo, 1988). These works investigated the gravitational stability of a two-phase system where the components interact primarily through their shared gravitational potential. The two fluids can be classified according to their relative abundance and "dynamical temperature." In galactic discs, stars represent the dominant and "hot" component, characterised by high velocity dispersion, while the gas is less abundant and "cold". Conversely, in protoplanetary discs, the roles are reversed: the gas is the dominant, hot component, while the dust is the less abundant, cold component. While dust is often treated as a collisionless fluid in these environments, for the purpose of gravitational stability it can be assigned an effective velocity dispersion $c_d$ which operates analogously to a sound speed. Bertin and Romeo (1988) demonstrated that gravitational instability can be driven by the cold component provided it is sufficiently abundant and cold. Specifically, the onset of this regime requires that

$$\frac{\Sigma_{\rm cold}}{\Sigma_{\rm hot}} > \frac{c_{\rm cold}}{c_{\rm hot}}. \tag{13}$$

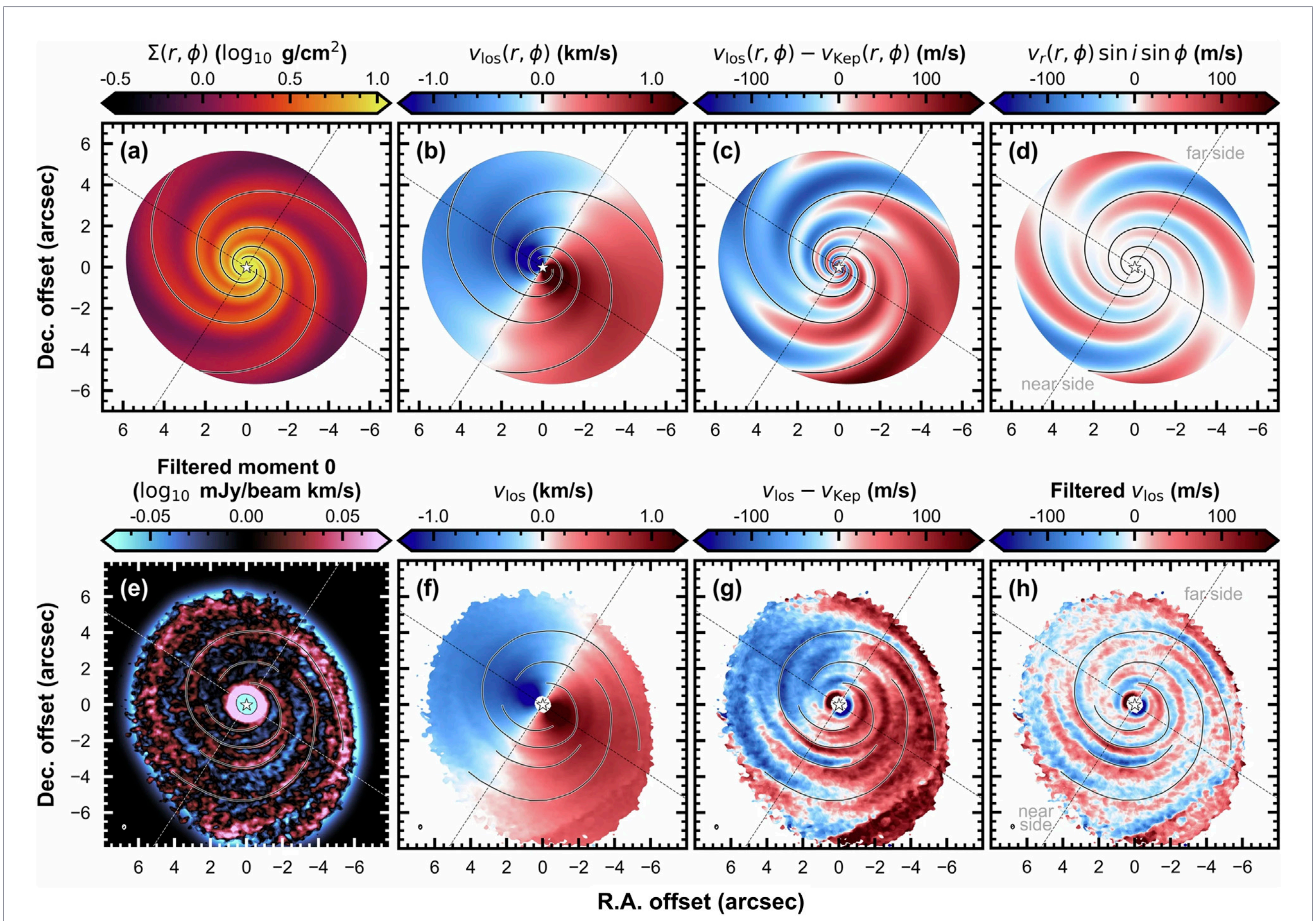

FIGURE 3
Comparison between analytical velocity field of a gravitational unstable disc (top panels, **a-d**) and Synthetic ALMA $^{13}CO$ observations of the 3D SPH GI disc simulation (bottom panels, **e-h**). $v_{los}$ corresponds to the line of sight velocity, $v_{kep}$ to the best fit Keplerian velocity and $v_r$ to the radial velocity. Readapted from Speedie et al. (2024). Global kinematic perturbations are visible in the observed velocity field, following the position of the spiral arms. The great agreement between the analytical model and the hydrodynamical simulation shows that the linear theory is well suited to describe the kinematics of gravitational unstable discs.

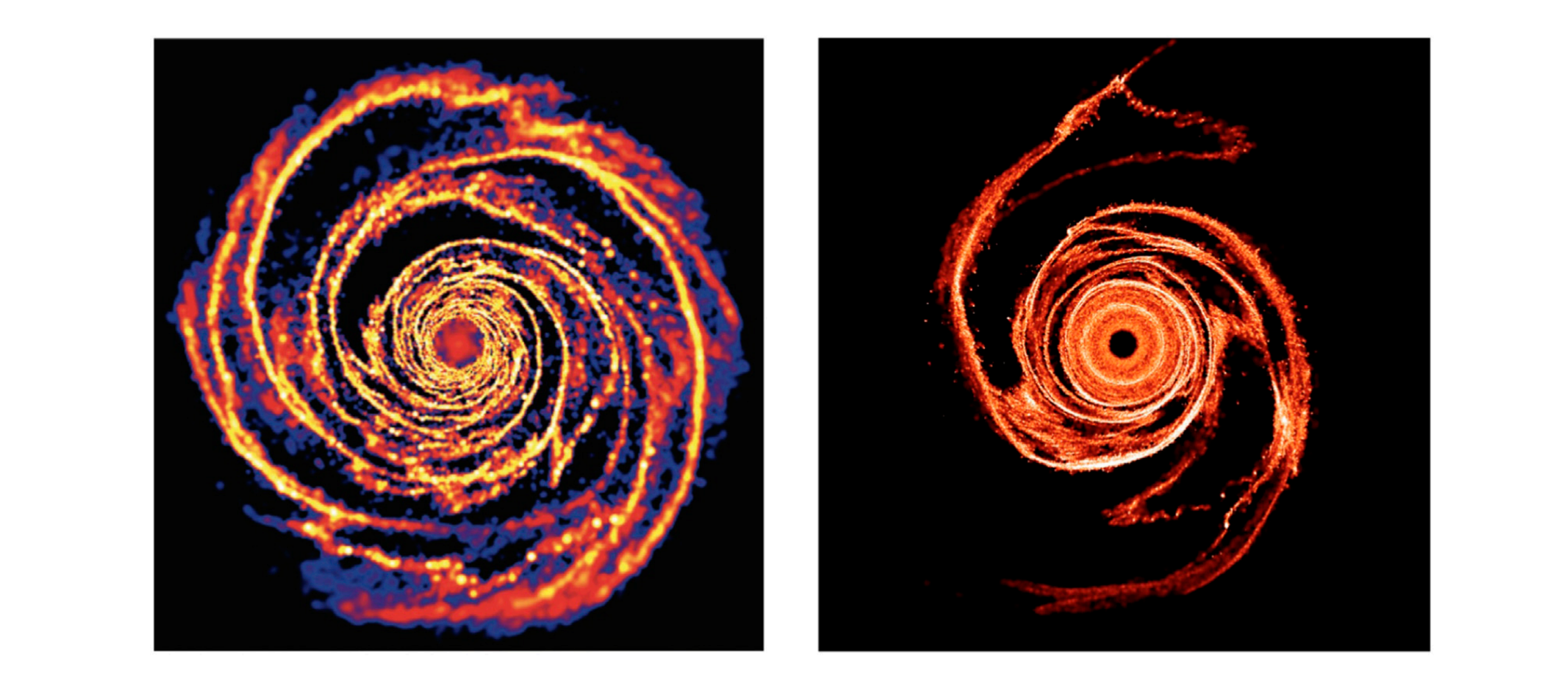

FIGURE 4
Dust surface density in 3D SPH global simulations (left panel Rice et al., 2004, right panel Rowther et al., 2024). In both panels, dust clumping is visible.

Longarini et al. (2023b) revisited this scenario by incorporating the role of the drag force, parameterised by the Stokes number (St). They found that for small Stokes numbers (i.e., strongly coupled particles), the destabilising effect of the solid component is negligible. However, as the Stokes number increases and the coupling weakens, the dust component contribution to the instability becomes increasingly significant. While $\epsilon = \Sigma_{\rm cold}/\Sigma_{\rm hot}$, $\xi = (c_{\rm cold}/c_{\rm hot})^2$, and St are treated as independent variables for the purpose of stability analysis, they are interdependent in a physical disc: the dust concentration typically peaks near $\mathrm{St} \approx 1$, and the velocity dispersion is itself a function of the gas–dust coupling. Furthermore, the authors derived the Jeans mass for these dust-driven perturbations. Due to the significantly lower dynamical temperature of the dust compared to the gas, the resulting Jeans mass is considerably smaller than in the gas-only fragmentation scenario, falling within the range of $1\text{–}10\,M_\oplus$, making this mechanism a viable way to create solid cores in young discs. These numerical models of dust dynamics in GI discs include dust backreaction. The dust particles are collisionless and inviscid, meaning that any diffusive mechanism arises solely from aerodynamic coupling with the gas.

These theoretical models highlight the critical role of the dust particles' kinematics. From a physical perspective, the interaction between gas spiral arms and the solid component induces relative motions among dust grains. These motions can be categorised into two distinct quantities: velocity dispersion and collisional velocity. The velocity dispersion (characterised by the standard deviation of particle relative velocities at a given location) represents a macroscopic pseudo-thermal property of the dust layer. This former quantity is what defines the effective sound speed $c_d$, which provides the pressure-like support against gravitational collapse discussed in the two-fluid stability analysis. Conversely, the collision velocity is defined on a pair-wise basis as the relative speed between two individual grains; it is the primary factor governing whether particles will grow through coagulation or undergo fragmentation upon impact.

The magnitude of $c_{\rm d}$ is regulated by the level of gravito-turbulence, set by the cooling timescale $\beta_{cool}$ (where faster cooling leads to higher excitation Booth and Clarke, 2016), and the Stokes number. Typically, $c_{\rm d}$ mirrors the gas turbulence for $St \ll 1$. It then has a minimum for $St \in [0.1, 1]$ before increasing for decoupled particles ($St > 1$) due to stochastic gravitational "kicks" from the spiral arms (Walmswell et al., 2013; Booth and Clarke, 2016; Riols et al., 2020). This velocity dispersion is the primary factor governing the stability of the solid component. Specifically, the criterion (Equation 13) defines the stability of the dust layer through the interplay between the relative dispersion and the local dust-to-gas ratio. There is a growing consensus within the community, supported by diverse numerical approaches, that dust grains with Stokes numbers in the range $St \simeq 0.1 - 1$ are the most effective candidates for rapid planetary cores formation. Indeed, particles in that range undergo the strongest concentration while maintaining a low velocity dispersion. This unique combination allows the solid component to satisfy the dust-driven GI criterion, overcoming the "dynamical heating" of the disc and leading to the direct gravitational collapse into planetary cores (Longarini et al., 2023a; Rowther et al., 2024, right panel of Figure 4).

To properly assess the conditions to activate dust collapse, recent works have employed local *shearing box* simulations, which allow for the high resolution necessary to capture the small-scale dynamics of the dust layer. Baehr et al. (2022) demonstrated that, in the presence of gravito-turbulence, dust grains can reach densities high enough to trigger collapse even when the gas itself is stable, confirming the analytical picture. Specifically, Baehr et al. (2022) and Baehr et al. (2025) identified a lower limit for the Stokes number required for collapse, finding that while $St \simeq 1$ particles are the most prone to fragmentation into clumps, the mechanism remains viable down to $St \simeq 0.05 - 0.1$. Below this threshold, the aerodynamic coupling is too strong, and the "dynamical heating" from the gas turbulence prevents the dust from settling into a sufficiently thin and dense layer to satisfy the condition in Equation 13. Furthermore, Rice et al. (2025) demonstrated that the success of the hybrid scenario depends on the dust size distribution. Direct core formation only occurs if a significant mass fraction is concentrated in the range $St > 0.1$. Conversely, a high abundance of small grains ($St < 0.1$) can inhibit collapse by increasing the effective velocity dispersion and preventing the dust from reaching critical densities. This underlines the importance of the dust growth process, to allow gravitational collapse to happen.

The viability of this hybrid scenario fundamentally depends on whether dust grains can grow to the required sizes ($St \gtrsim 0.1$) within the short timescales of the GI-active phase. For a typical disc mass of $M_{disc} \sim 0.1\,M_\odot$, a Stokes number of 0.1 corresponds to a physical grain size of roughly 0.5 mm, a significant growth milestone for such a young system. We underline that the grain size estimate strongly depends on the radial distance from the central star. Standard dust growth models in smooth discs struggle to reach these sizes so rapidly (Birnstiel, 2024), as they are further hindered by the radial drift barrier, especially in the outer disc. However, gas spiral arms act as pressure traps that not only halt radial drift but also create high-density environments where grain growth could be significantly accelerated. To determine if this acceleration leads to successful growth or to a fragmentation bottleneck, it is necessary to investigate the collisional velocity between grains.

The collisional velocity, $v_{coll}$, defined on a pair-wise basis as the relative speed between two individual grains, is the primary factor governing whether particles grow through coagulation or undergo fragmentation upon impact. In gravito-turbulent discs, the dynamics of collisions are significantly more violent than in laminar flow, creating a potential bottleneck for the hybrid planet formation scenario.

Booth and Clarke (2016) provided the first systematic characterisation of the collisional velocity in the context of GI,

demonstrating that it scales as $v_{coll} \propto St^{1/2}$. This scaling is physically rooted in the nature of the gas dynamics; as shown by Booth and Clarke (2019) through spectral analysis, GI-driven turbulence follows a Kolmogorov-like cascade. Since dust grains are accelerated by the turbulent gas velocity field, their relative motions reflect the properties of the turbulent cascade, leading to the square-root dependence on the Stokes number. However, a crucial distinction must be made between *mono-disperse* and *multi-disperse* dust populations. In a mono-disperse distribution, where all grains share the same Stokes number, relative velocities are the smallest, because all the particles react identically to gas fluctuations. Conversely, in a multi-disperse environment, the difference in aerodynamic coupling between grains of different sizes significantly increases their relative motion. This makes the fragmentation barrier ($v_{frag}$) a big obstacle for grain growth. For silicate-dominated grains, which fragment at velocities of approximately $1\,m/s$ (Gundlach and Blum, 2015; Musiolik and Wurm, 2019), GI-induced motions can easily exceed the threshold at Stokes numbers as low as $St \sim 10^{-2}$. This suggests that reaching the critical $St \simeq 0.1-1$ regime required for dust-driven GI might be extremely challenging for rocky material. On the contrary, for icy aggregates with higher fragmentation thresholds ($10-30\,m/s$, Wada et al., 2013; Musiolik, 2021), growth remains a viable pathway, particularly within the high-density environment of spiral arms where the concentration of solids can significantly enhance the growth rate.

Even though self-gravity is not involved, it is worth comparing this hybrid planetary cores formation scenario with the streaming instability (SI), as both mechanisms ultimately lead to the direct gravitational collapse of concentrated solids into planetesimals or planetary cores. Yet, they differ substantially in their physical origin and in the disc conditions under which they operate. The SI is fundamentally an aerodynamic instability, driven by the mutual drag force between gas and dust: local overdensities of solids exert a backreaction on the gas that reduces its effective sub-Keplerian velocity, which in turn reduces the relative drift velocity of nearby particles, reinforcing the clumping of solid particles. Crucially, this mechanism does not rely on the self-gravity of the gas disc, and it is most efficient when the local dust-to-gas ratio approaches or exceeds unity, a condition often achieved through independent concentration processes. Dust-driven GI, by contrast, is intrinsically related to the large-scale self-gravity of a massive disc: the concentration of solids is a direct consequence of the spiral structure induced by the gas gravitational instability, and the subsequent collapse of the dust layer is governed by its own effective Jeans criterion. In this sense, the two mechanisms are in some respects complementary, rather than competing: dust-driven GI is expected to operate preferentially at very early times, when the disc is still massive enough to be self-gravitating, whereas the SI is likely to become relevant later in the disc lifetime, once the disc has evolved towards lower masses and dust has had time to settle and grow in a smoother disc.

Taken together, these results establish gravitational instability as a viable path to form planets, not through direct gas fragmentation, but by creating the physical conditions for the gravitational collapse of the solid component. This offers a rapid pathway to form planetary cores in the earliest stages of disc evolution.

## 3.3 GI and thermodynamics

The non-linear evolution of gravitational instability and its ability to lead to fragmentation are fundamentally governed by the thermodynamics of the disc. Historically, the most common approach to model it has been the $\beta$–cooling prescription (Gammie, 2001), where the cooling timescale is considered proportional to the dynamical timescale, with $\beta$ the proportionality factor. While numerically efficient and useful to identify the basic physics of gravitoturbulence, this is certainly not realistic. Indeed, this method neglects the complex physics of radiative cooling and shock heating, which in real discs depends strongly on local properties such as density, temperature, and, most critically, the optical depth $\tau$, which varies significantly across the radial extent of the disc.

Firstly, Stamatellos et al. (2007) tackled this problem in 3D global SPH simulations, introducing more realistic thermodynamics through a local radiative cooling and heating approximation. By estimating the optical depth at every particle location in the simulation, according to the local disc properties, this method allowed for the first time to incorporate a physically motivated energy equation into SPH simulations without the prohibitive cost of full radiative transfer. Despite its success, this method relies on a simplified estimate of the optical depth based on the local column density, which may not fully capture the complex 3D geometry of radiation field in a highly perturbed disc.

As a direct evolution of the Stamatellos et al. (2007) approach, Young et al. (2024) introduced a refined thermodynamic scheme, building upon the work of Lombardi et al. (2015), to improve the accuracy of radiative cooling in global hydrodynamical simulations. Their approach is based on a polytropic cooling approximation, under the assumption that each gas parcel resides within a "spherical pseudo-cloud" to estimate the local optical depth. This hybrid framework overcomes the limitations of earlier models by providing higher accuracy in optically thick environments, where the original Stamatellos et al. method often faced challenges. In this framework, heating is treated as a prescribed mechanical response to the gas dynamics. Rather than being derived from a radiation field, the thermal energy input is calculated locally through artificial viscosity, which converts kinetic energy into heat at shock fronts, and adiabatic compression ($PdV$ work). This approach ensures numerical efficiency by coupling the thermodynamics directly to the hydrodynamic variables, though it bypasses radiative feedback mechanisms.

A more fundamental approach is adopted by Xu et al. (2025a), Xu et al. (2025b), who transition from approximate prescriptions to solving the full radiation transport equation. These studies employ grid-based global simulations using the *Athena++* code to resolve the interaction between gas and radiation along 32 discrete angles. In this high-fidelity framework, the thermodynamics is not prescribed but is instead an emergent property of the system. Rather than relying on local shock-heating recipes, the thermal state is self-consistently determined by solving the radiative transfer equation throughout the disc. This approach captures non-local feedback mechanisms, such as radiative "self-heating", where energy released in one region can travel and warm distant areas, a process entirely neglected by previous methods. While this represents the current "Gold Standard" for

TABLE 1 Comparison of different radiative and thermodynamic implementations in recent 3D global simulations of self-gravitating discs.

| Reference | Radiative technique | Heat source | Computational cost |
|---|---|---|---|
| Young et al. (2024) | Local recipes | Internal (shocks) | Low |
| Xu et al. (2025a), Xu et al. (2025b) | Full transport | Internal (emergent) | Very high |
| Rowther et al. (2024) | Monte carlo (MCRT) | Stellar irradiation | Very high |

References: 1 Young et al. (2024); 2 Xu et al. (2025a), Xu et al. (2025b); 3 Rowther et al. (2024).

capturing the internal physics of self-gravitating discs, it comes at a formidable computational cost. Solving the transport equation along dozens of angles requires orders of magnitude more resources than approximate methods, currently limiting these simulations to shorter evolution.

A significant novelty has been introduced by Rowther et al. (2024), who moved the focus from the disc internal energy balance to the influence of the stellar irradiation. While previous models concentrate on how the disc manages heat generated internally, this study employs the SPH code phantom coupled with the radiative transfer code mcfost to perform live Monte-Carlo Radiative Transfer (MCRT) simulations. This approach allows for the fundamental inclusion of stellar irradiation. In this scenario, the disc thermodynamics is no longer dictated solely by GI self-regulation, but is instead dominated by the radiative flux from the central star. This irradiation imposes a “thermal floor” that is often significantly higher than the heat generated locally by GI-induced shocks. This external heating breaks the feedback loop required for gravitational instability to self-regulate, suppressing or entirely preventing the formation of spiral arms in the disc. This approach comes at an incredibly high computational cost. A summary of the three aforementioned thermodynamic implementations is given in Table 1.

Before (Rowther et al., 2024), the role of stellar irradiation have been investigated extensively. Rice et al. (2011) were the first to systematically investigate the fragmentation boundary in irradiated self-gravitating discs, showing that irradiation weakens disc self-gravity and that the critical $\alpha$ for fragmentation decreases as irradiation increases, meaning that strongly irradiated discs require more vigorous cooling to fragment. Building on this, Forgan and Rice (2011) extended the Jeans mass formalism to irradiated discs, demonstrating that the addition of stellar irradiation raises the equilibrium disc mass required for self-gravity to operate at a given accretion rate, and shifts the region of parameter space where fragmentation is expected to occur. Building upon those works, Haworth et al. (2020) and Cadman et al. (2020) extended the investigation of irradiation effects to explore how the susceptibility of a disc to gravitational instability and fragmentation depends on the mass of the host star. They found that discs become increasingly prone to the effects of fragmentation as the stellar mass increases. This happens because, at fixed Toomre $Q$ and disc-to-star mass ratio, discs around more massive (and hence more luminous) stars are hotter, requiring higher heating and cooling rates to sustain energy balance. Since the effective $\alpha_{GI}$ scales directly with this cooling rate, this results in a higher $\alpha_{GI}$ or discs around high-mass stars. In the optically thin limit, stellar irradiation dramatically stabilises discs around low-mass stars. However, even in this strongly irradiated regime, discs around $\sim 2\,M_\odot$ stars remain prone to fragmentation, preferentially producing wide-orbit giant planets and brown dwarfs. A complementary consequence of this result is that low-mass stars may support high disc-to-star mass ratios without triggering GI, providing large mass reservoirs for planet formation.

A complementary perspective on the role of stellar irradiation is offered by Leedham et al. (2025), who address a more fundamental question: how does the mathematical implementation of irradiation heating affect the outcome of GI simulations? Working within the framework of 2D local shearing-sheet simulations with $\beta$-cooling, they compare two prescriptions for the irradiation heating term, a constant rate per unit mass versus per unit area of the disc. This distinction, rooted in whether the disc is optically thin or thick to the incident stellar flux, has profound consequences for the disc stability. In the former case, overdense regions are preferentially heated, possibly suppressing GI, in line with the earlier findings of Rice et al. (2011). In the latter case, gravitoturbulence persists even at high irradiation levels, with the disc reaching a self-regulated state at values of $Q \sim 10$, well above the classical threshold. Crucially, fragmentation remains possible in this highly irradiated, optically thick regime, challenging the conventional expectation that strongly irradiated outer discs are stable against collapse.

A related but distinct aspect of irradiation heating concerns its effect on the linear stability of the disc. Lin and Kratter (2016) performed a stability analysis of discs subject to external heating, demonstrating the existence of growing modes at scales below the thermal Jeans length ($kH > 1$), a regime not captured by standard GI theory. Löhnert et al. (2020) confirmed these growth rates through 2D hydrodynamical simulations of externally irradiated discs, finding significant mode growth even when $Q > 1$. Crucially, in the saturated state, this instability drives angular momentum transport corresponding to an effective $\alpha \sim 10^{-3}$, even for Toomre parameters as high as $Q \sim 10$.

Taken together, these studies underscore the central role of thermodynamics in shaping the outcome of gravitational instability. While the $\beta$–cooling prescription remains a simplified approximation, it continues to provide an invaluable conceptual framework, isolating the fundamental physics of gravito-turbulence before the added complexity of realistic radiative transfer is introduced.

## 3.4 GI and planets: migration, ejection and interaction with spiral arms

Gravitational instability has the potential to create embedded companions, either through the direct collapse of the dust component (Rice et al., 2006; Longarini et al., 2023a), forming super-Earth mass objects, or through gas fragmentation (Boss, 1997), forming Jupiter-like planets and brown dwarfs. However, the fate of these bodies within a massive, strongly perturbed disc remains an open question.

Planet migration in self-gravitating discs is fundamentally different from its counterpart in standard, non-self-gravitating discs. In massive discs where GI is active, the gravitational stresses associated with spiral density waves generate strong, time-varying torques on any embedded companion. These stochastic torques, arising from the perturbed density field of the disc, can significantly alter the migration rate and direction compared to the smooth, laminar predictions of classical Type I and Type II migration theory (Paardekooper et al., 2023; Benitez-Llambay, 2026). The first systematic numerical studies of planet migration in self-gravitating discs were carried out by Baruteau et al. (2011) and Malik et al. (2015), using two- and three-dimensional simulations with a globally constant cooling parameter $\beta$. Their results showed that planets of a wide range of masses migrate inward extremely rapidly, with migration rates well in excess of classical Type I predictions. More strikingly, gap opening, the standard mechanism by which massive planets transition to the slower Type II migration regime, was found to be severely inhibited, even for substellar companions as massive as $30M_{jup}$.

A key limitation of these studies is the use of a globally constant $\beta$. This prescription forces the disc to be gravitationally unstable everywhere, generating spiral structure throughout its entire radial extent. This is, however, not expected from physically motivated models of protostellar discs, where the cooling efficiency decreases strongly toward the inner regions due to the increasing optical depth. As a consequence, real self-gravitating discs are expected to be gravitationally unstable only in their outer parts, with the inner disc remaining in a stable, hotter state (Rafikov, 2005; Clarke, 2009; Kratter and Lodato, 2016). Rowther and Meru (2020) addressed this issue by implementing a radially varying cooling parameter which ensures that the cooling timescale in the inner disc is long enough to prevent gravitational instability, while the outer disc remains unstable and develops the spiral structures. In this setup, planets with different initial masses, ranging from $1M_{\oplus}$ to $5M_{jup}$ initially undergo strong Type I migration, consistent with Baruteau et al. (2011). However, when they reach the gravitationally stable inner region, migration significantly slows down, without requiring the planet to open a gap (Rowther and Meru, 2020). This happens because in the outer turbulent region, strong asymmetric density fluctuations produce a negative net torque on the planet, driving rapid inward migration. Conversely, in the inner, stable regions, the density field is much smoother and the corotation torque becomes more symmetric, reducing the net torque and allowing the planet to reach a quasi-stationary state. These results indicate that the thermodynamics of the disc plays a fundamental role in determining the migration rate of planets embedded in self-gravitating discs.

Nevertheless, the interaction between an embedded planet and a self-gravitating disc is not unidirectional. While the disc torques drive planetary migration, the planet itself can significantly alter the structure of the disc. Rowther et al. (2020) showed that an embedded giant planet can suppress the spiral structure driven by gravitational instability. As the planet interacts with the disc, the spiral density waves it generates propagate radially and create shocks, depositing energy and heating in the surrounding gas. This increase in sound speed raises the Toomre parameter above the instability threshold, causing the disc to become gravitationally stable even though its mass has not changed. The net result is a disc that, despite being massive enough to be self-gravitating in the absence of a planet, appears largely axisymmetric with a ring and gap structure in continuum observations. This has a profound implication for the interpretation of ALMA observations, since spiral arms are often considered the smoking gun of gravitational instability (Rowther et al., 2023). A massive disc hosting a giant planet may therefore appear gravitationally stable, making the two scenarios observationally degenerate, as shown in figure Figure 5.

Planet migration is not the only physical process occurring to bodies formed or embedded in self-gravitating discs. Indeed, the fate of the planetary objects is governed by a competition between several timescales: the gaseous collapse timescale, the grain growth and settling timescales, the migration timescale, and the mass loss timescale due to tidal stripping. This framework, known as *tidal downsizing* (Nayakshin, 2010), predicts a wide zoo of final objects depending on how these timescales are ordered. Forgan and Rice (2013), Forgan et al. (2018a) explored this scenario through a population synthesis model that incorporates fragment-fragment gravitational interactions during the disc phase and tidal downsizing (see also the recent DIPSY model by Schib et al., 2025). Their key finding is that dynamical scattering between fragments dominates over migration in shaping the final population: typical systems produced by GI consist of one or two massive objects at large separations, with a small fraction of terrestrial-type bodies and a significant population of ejected free-floating planetary mass objects (Nayakshin et al., 2026). In addition, the chemical composition of GI fragments is also expected to deviate significantly from the bulk disc material. Ilee et al. (2017) showed that fragments can undergo a "cold start" phase after formation, during which temperatures are not high enough to fully desorb volatile ices from grain surfaces. If dust sedimentation occurs on a comparable timescale, solid cores can form with a C/O ratio that differs from that of the local disc, meaning that the atmospheric composition of planets formed via GI does not necessarily reflect the bulk composition of the disc from which they formed.

## 3.5 GI and infall

Star and planet formation is the result of a complex, multi-scale process involving gravitational collapse and angular momentum transport. The transport of material from the large scales of molecular cloud cores down to the smaller scales of protostellar discs is a fundamental aspect of the earliest phases of disc assembly. In this context, the disc cannot be treated as an isolated system; rather, its evolution is strongly linked to the interaction with the parental envelope through mass transfer. Consequently, it is of great interest to understand the regime where gravitational instability is triggered and sustained by infall, rather than being purely regulated by cooling. This "infall-driven" GI is expected to be the dominant mechanism during the Class 0 and I phases, where the rapid mass infall rates can drive the disc toward instability long before secular cooling processes become significant. The connection between infall and gravitational instability was first explored in a series of numerical studies by Vorobyov and Basu (2005a), Vorobyov and Basu (2005b), Vorobyov and Basu (2007); Kratter et al. (2008), (2010), who modelled the self-consistent collapse of rotating molecular cloud cores into protostellar discs.

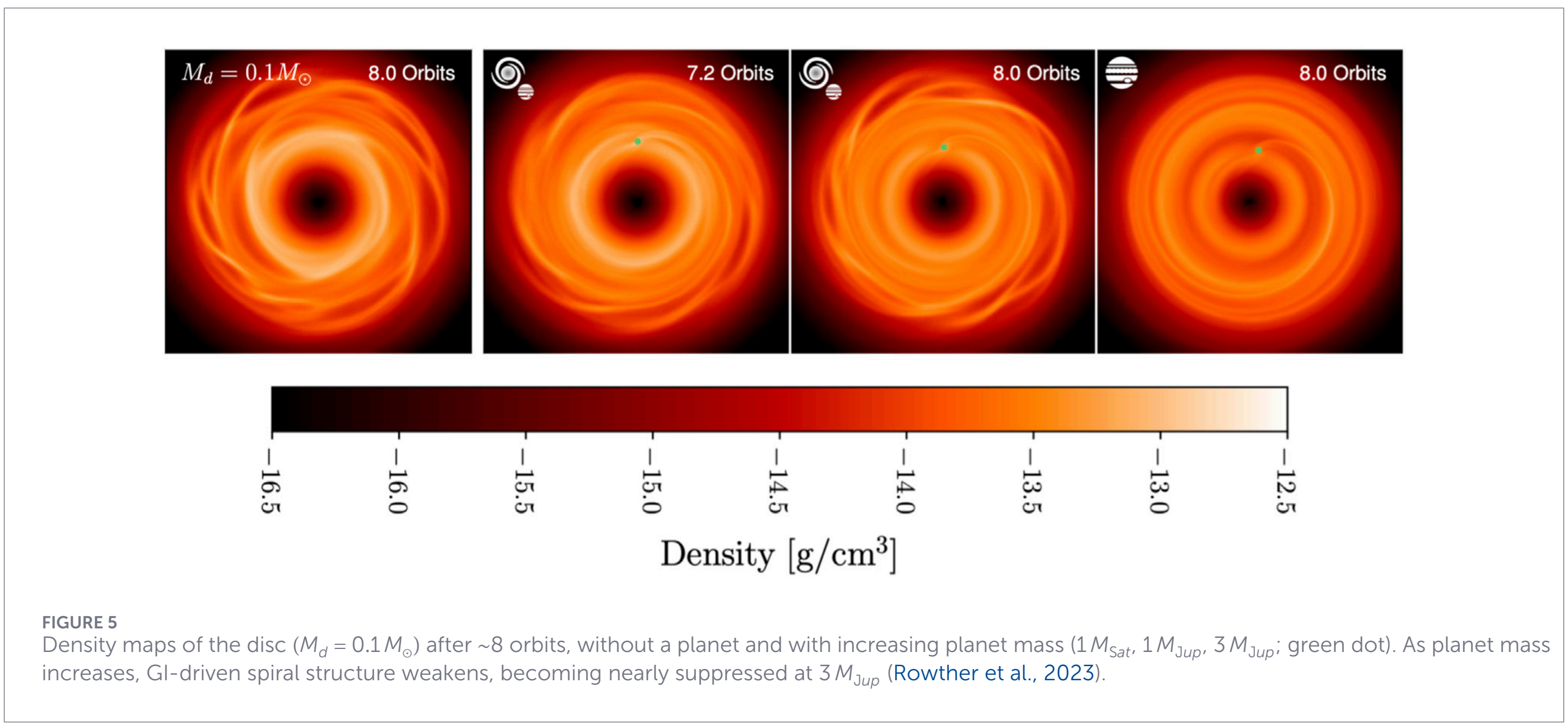

FIGURE 5
Density maps of the disc ($M_d = 0.1\,M_\odot$) after ~8 orbits, without a planet and with increasing planet mass ($1\,M_{Sat}$, $1\,M_{Jup}$, $3\,M_{Jup}$; green dot). As planet mass increases, GI-driven spiral structure weakens, becoming nearly suppressed at $3\,M_{Jup}$ (Rowther et al., 2023).

These simulations showed that continuous mass loading from the envelope drives the disc into recurrent gravitational instability, leading to fragmentation and the inward migration of dense clumps, which generate episodic accretion bursts analogous to those observed in FU Ori-type objects.

Significant efforts have been made in the last few years to study the formation of protoplanetary discs in simulations of molecular cloud collapse. Kuffmeier et al. (2018) used large-scale "zoom-in" simulations to demonstrate that discs formed during the collapse of molecular cloud cores are naturally prone to gravitational instability. Their results show that the primordial infall of material from the envelope frequently drives the disc into an unstable regime, leading to episodic accretion bursts. These bursts are a fundamental mechanism for processing the rapidly accumulating mass, providing a possible physical explanation for the observed variability in young protostellar objects. This picture is further refined by the recent work of Lebreuilly et al. (2024), who investigated the early "build-up" phase of protoplanetary discs. Their models indicate that during the first $10^4$ years of disc assembly, these systems are exceptionally massive, often reaching a disc-to-star mass ratio $M_d/M_\star \sim 1$. In this extreme regime, discs are inherently prone to gravitational instability and potentially to early fragmentation. However, the authors demonstrate that radiative feedback plays a crucial role in regulating this process. By accounting for the heat generated by accretion onto the protostar (namely, accretion luminosity), the disc temperature increases, acting as a powerful stabiliser that significantly reduces the occurrence of fragmentation compared to simpler barotropic models.

Building on this framework, Longarini et al. (2025b), following the works of Kratter et al. (2008), (2010), Zhu et al. (2012), formalised the concept of "infall-driven" gravitational instability, highlighting a fundamental self-regulation mechanism. Unlike isolated cooling-driven discs, where the stability is primarily governed by the cooling timescale $\beta_{cool}$, infall-driven discs are regulated by the rate of mass transferred from the envelope. In this scenario, the disc self-adjusts to a state of marginal stability ($Q \approx 1$) by balancing the incoming mass with GI-induced angular momentum transport, redistributing it between the central star and the outer disc. The resulting effective viscosity $\alpha$ and the amplitude of the spiral structures are thus dictated by the infall rate rather than local thermodynamics. We underline that there is an important underlying hypothesis, that is the locally isothermal equation of state. We expect that in a scenario where also temperature can evolve, the temperature regulation should also play a role in the infall-driven scenario. This self-regulation ensures that the disc can efficiently process the material injected from the large-scale environment, maintaining its stability throughout the high-mass-accretion phase of Class 0 and I sources.

Infall from the environment is known to produce misalignments and warps in protostellar discs (Kuffmeier et al., 2021). Rowther et al. (2022) studied the interplay between warps and gravitational instability, showing that as a warp propagates through a self-gravitating disc it heats the gas, raising the Toomre parameter above the instability threshold and suppressing the spiral structure entirely. The disc consequently appears axisymmetric in continuum observations, providing yet another mechanism, alongside planet-disc interactions, by which the signatures of GI can be hidden in young, massive discs.

These results signal a growing consensus that the environment surrounding a young disc cannot be ignored. Infall and gravitational instability are not independent processes, but act together in shaping the early disc structure and the condition for planet formation.

# 4 Observational advances on GI

## 4.1 Dynamical masses

Throughout the paper, we stressed how important is the disc mass to assess whether the disc is gravitationally unstable. However, Weighing protoplanetary discs is challenging (Miotello et al., 2023) because molecular hydrogen ($H_2$), that is the main gaseous component, is invisible in such cold environments. Consequently, alternative tracers must be used to probe the disc mass. Common gas tracers include CO isotopologues (Miotello et al., 2017), which are chemically simple and among the most abundant molecules in discs. Another option is HD (e.g., Bergin et al., 2013), the closest analogue to $H_2$, which provides a more direct probe of the total mass but the main transition is not currently covered by any operating facilities. Dust can also be used to infer the disc mass, under the assumption of optically thin emission and an assumed gas-to-dust ratio (e.g., 100 as in the ISM). However, all these methods rely on hypothesis regarding disc composition, optical depth, and tracer-to-$H_2$-conversion factors, which are often uncertain.

A powerful and more direct approach, free from these assumptions, has been developed in the last 5 years, and consists in estimating the disc mass dynamically. This method consists in thoroughly modelling the rotation curve of the disc (analogously to what done in galactic discs), including the self-gravitating contribution, in order to constrain its mass. Indeed, a gas particle at a distance $R$ from the central star will experience the Keplerian gravitational force of the star, and also the gravitational pull of the disc material within its orbit. In addition, the pressure gradient must be considered, slowing down the azimuthal gas rotation.

The first application of the model was done for the Elias 2–27 disc in Veronesi et al. (2021) and later refined in a series of papers. Here we present the latest model of the rotation curve of a protoplanetary disc, presented in Martire et al. (2024), as a generalisation of the vertically isothermal model presented in Lodato et al. (2023) and Veronesi et al. (2021). This method has been used to constrain dynamical disc masses of discs within MAPS and exoALMA sample (Martire et al., 2024; Longarini et al., 2025a), and benchmarked against numerical models (Veronesi et al., 2024; Andrews et al., 2024). According to the centrifugal balance equation, the rotation curve of a protoplanetary disc can be written as

$$v_\phi^2 = R\frac{\partial \Phi_\star}{\partial R} + R\frac{\partial \Phi_d}{\partial R} + \frac{R}{\rho}\frac{\partial P}{\partial R}, \qquad (14)$$

where the first contribution on the right hand side is the stellar one, namely, the Keplerian rotation at a given position $(R, z)$ in the disc, where $\Phi_\star$ is the stellar (Keplerian) gravitational potential. The second contribution corresponds to the self-gravitating contribution of the disc, where $\Phi_d$ is the gravitational potential of the disc, proportional to the disc mass. Finally, the third contribution is the pressure gradient, where $P$ is the gas pressure.

The self-gravitating contribution can be computed by solving the Poisson equation for a thin disc, as done in Bertin and Lodato (1999). This contribution is directly proportional to the disc mass, making it possible to constrain it with high-resolution kinematic data, when the disc is massive enough. The self-gravitating contribution is degenerate with the pressure gradient, that should be thoroughly described in order to provide a solid disc mass estimate. In the latest rotation curve model, introduced by Martire et al. (2024) and used in Pezzotta et al. (2025), Longarini et al. (2025a), the disc is assumed to be vertically stratified in temperature. This stratification is described in terms of a function $f$ such that $T(R,z) = T_{mid}(R) f(R,z)$. The temperature stratification induces a density stratification, because of the hydrostatic equilibrium, that can be described through a function $g$ such that $\rho(R,z) = \rho_{mid}(R) g(R,z)$. The functions $f$ and $g$ are related through the hydrostatic equilibrium as shown in Martire et al. (2024) (in the limit of vertically isothermal discs, $f(R,z) = 1$, and for $z \ll R$, $g(R,z)$ reduces to the usual Gaussian profile often adopted for Keplerian discs). Finally, a surface density profile must be specified to compute both the self-gravitating and the pressure gradient contribution. So far, a self-similar solution for the surface density has always been employed, depending on two parameters, namely, the disc mass and the disc scale radius. Within those assumptions, the rotation curve of a vertically stratified protoplanetary disc is

$$v_\phi^2(R,z) = v_\mathrm{k}^2 \left\{ \underbrace{\left[1 + \left(\frac{z}{R}\right)^2\right]^{-3/2}}_{\text{Height correction}} - \underbrace{\left[\gamma' + (2-\gamma)\left(\frac{R}{R_c}\right)^{2-\gamma}\right.}_{\text{Pressure gradient at midplane}} - \underbrace{\left.\frac{d\log(fg)}{d\log R}\right]}_{\text{Vertical stratification}} \left(\frac{H}{R}\right)^2_{\mathrm{mid}} f(R,z) \right\} + \underbrace{+\, v_d^2(R,Z)}_{\text{Self-gravity}} \qquad (15)$$

where $v_k^2 = GM_\star/R$ $\gamma$ is the power law coefficient of the surface density profile, $\gamma' = \gamma + (3 + q_{\mathrm{mid}})/2$, where $q_{\mathrm{mid}}$ is the power law index of the temperature profile at the disc midplane $T_{\mathrm{mid}} \propto R^{-q_{\mathrm{mid}}}$. The self-gravitating contribution $v_d$ has the following form (Bertin and Lodato, 1999)

$$v_d^2 = G\int_0^\infty \left[K(k) - \frac{1}{4}\left(\frac{k^2}{1-k^2}\right) \times \left(\frac{R'}{R} - \frac{R}{R'} + \frac{z^2}{RR'}\right)E(k)\right]\sqrt{\frac{R'}{R}}k\Sigma(R')\,dR', \qquad (16)$$

where $K$ and $E$ are the elliptic integral of the first kind and $k^2 = 4R'/((R+R')^2 + z^2)$. Equation 15 relies on several parameters that characterise the disc. These can be divided into two categories: structural parameters, which define the underlying disc model, and free parameters, which are typically treated as variables to be constrained through a fitting procedure against kinematic observations. The structural parameters are

- Height of the emitting layer for the chosen tracer, $z(R)$:
  When probing the rotation curve with a molecular tracer, such as CO isotopologues, the emission originates from an elevated surface known as the emitting layer. In the process of extracting rotational profiles, the height of this emitting layer can be retrieved as well with high precision (Izquierdo et al., 2025; Galloway-Sprietsma et al., 2025).
- Surface density slope, $\gamma$:
  Following the assumption of self-similar solution for the surface density, one needs to specify that is the surface density slope $\gamma$. So far, all the dynamical estimates have used $\gamma = 1$. Andrews et al. (2024) tested the robustness of the measurements against different value of $\gamma$, finding that it does not significantly impacts the dynamical mass estimate.
- Thermal structure, $f(R,z)$:
  The thermal structure determines the pressure gradient contribution, so a good estimate of it is crucial in getting a reliable disc mass. Indeed, the pressure gradient contribution is degenerate with the self-gravitating one. So far, 2D thermal structures have been determined through optically thick molecular tracers (Galloway-Sprietsma et al., 2025).

The free parameters that can be fitted to kinematic observations are

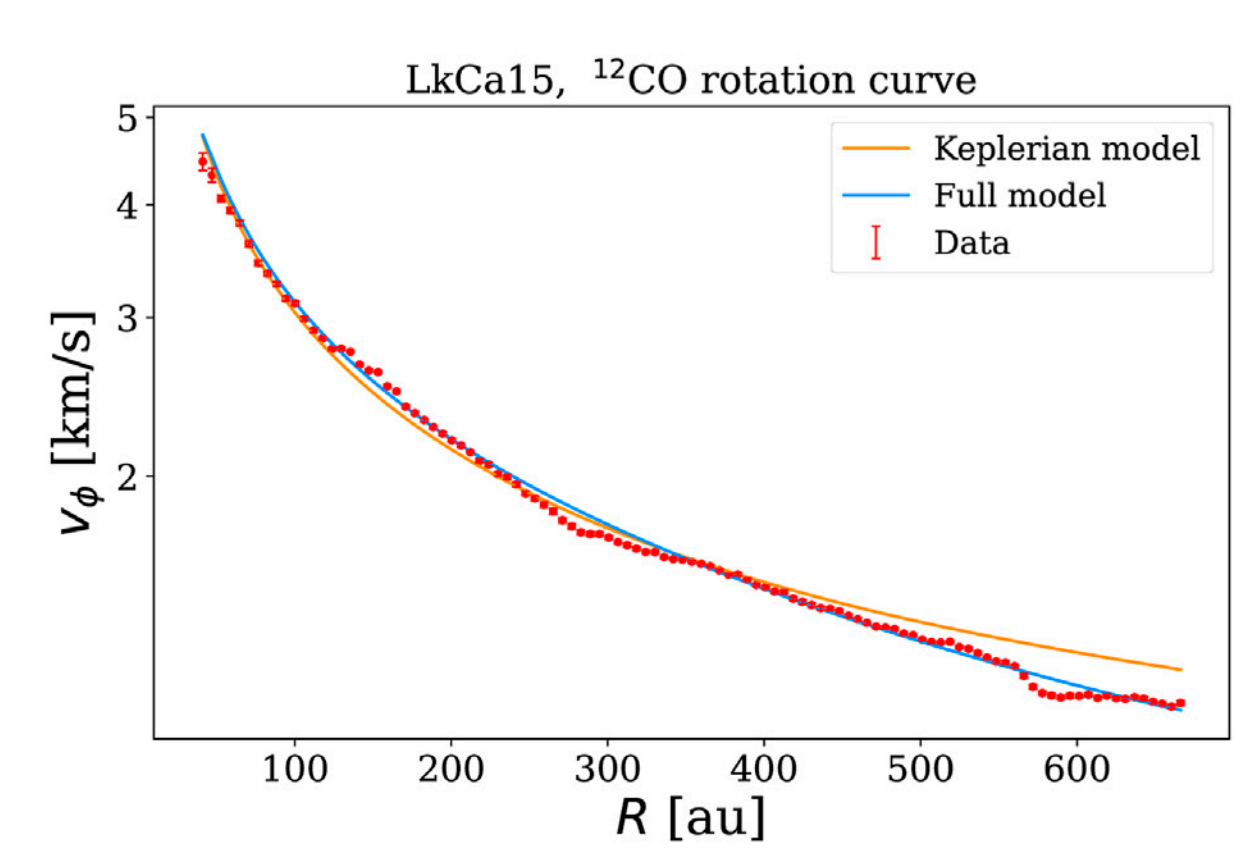

FIGURE 6
$^{12}$CO rotation curve of the protoplanetary disc LkCa15 (red dots) compared with a purely Keplerian model (orange line) and a full model incorporating pressure gradient and disc self-gravity contributions (blue line). The comparison shows that a simple Keplerian profile is insufficient to reproduce the observed rotation curve, highlighting the need to account for the disc gravitational potential. Adapted from Longarini et al. (2025a).

- Stellar mass, $M_\star$:
  The stellar mass determines the underlying Keplerian rotation profile, and appears inside the term $v_k$.
- Disc mass, $M_d$:
  The disc mass determines the strength of the self-gravitating contribution
- Disc scale radius, $R_c$:
  The disc scale radius determines the strength of the pressure gradient contribution.

An example of the fitting procedure is shown in figure Figure 6. This method has proven highly effective in providing dynamical mass estimates with relatively small uncertainties. Indeed, stellar masses are retrieved with a typical uncertainty of ~2%. It is important to emphasize that these stellar masses represent the most accurate estimates available to date, as all the different components contributing to the rotational profiles are taken into account. This marks a significant improvement compared to a simple Keplerian fit of the rotational data, which can result in discrepancies on the order of ~10 – 20% (Longarini et al., 2025a). As for disc masses, the typical uncertainty is of the order of ~20 – 30% (Veronesi et al., 2024; Andrews et al., 2024). It appears that the retrieval of the thermal structure is the dominant source of uncertainty for the disc mass, making it crucial to accurately characterise the 2D temperature structure. Pezzotta et al. (2025) employed seven molecular lines to retrieve the thermal structure of the HD 163296 disc, subsequently extracting the rotation curves for these same molecules to fit for the disc mass. Their findings demonstrate that increasing the number of molecular tracers leads to a more robust characterisation of the thermal structure, which in turn significantly reduces the uncertainty associated with the disc mass estimate.

Table 2 collects the 18 discs with available dynamical mass measurements. When comparing the masses of protoplanetary discs that can be probed with this dynamical method with standard continuum-based estimates, it is evident that the sample populates the very high end of both the disc and stellar mass distributions (Figure 7). Indeed, the dynamical mass estimates (orange dots) consistently correspond to the most massive systems within the broader Manara et al. (2023) PPVII sample (grey dots). Primarily, it is a physical limitation: the self-gravitating contribution becomes significantly detectable only in the most massive discs, where the disc-to-stellar mass ratio is high enough to induce measurable deviations from a purely Keplerian potential. Secondly, there is an observational limitation. The successful application of this method relies on high spatial and spectral resolutions. The target discs must be sufficiently extended and bright in molecular line emission to provide as many resolution elements as possible across the disc's surface. These requirements naturally favour large, gas-rich systems around intermediate/high-mass stars, which are precisely the objects located at the very high end of the standard populations.

Dynamical masses and thermal structures of protoplanetary discs provide a direct means to assess the onset of gravitational instability. Computing the Toomre parameter requires knowledge of three key quantities: the midplane temperature profile, the surface density profile, and the stellar mass. It is worth noting that all three are model dependent. The thermal structure is inferred under the assumption that the functional form of the two-dimensional temperature profile is known (Galloway-Sprietsma et al., 2025), with the midplane temperature extrapolated from this function, since molecular line observations probe the $^{12}$CO and $^{13}$CO emitting layers rather than the midplane directly. The surface density profile is prescribed as a self-similar solution, meaning that the dynamical estimates of the disc and stellar mass depend on this choice. These assumptions are physically well motivated, but their cumulative effect means that the resulting Toomre parameter values should be interpreted with appropriate caution.

Table 2 collects the minimum Toomre parameter for the discs with available dynamical mass measurements. It is remarkable that the systems with the lowest values of $Q$, namely, Elias 2–27, GM Aur, IM Lup, and WaOph 6, are precisely those showing spiral structures, either in dust continuum emission (IM Lup, WaOph 6), molecular line emission (GM Aur), or both (Elias 2–27). The apparent outlier is HD 97048: pezzotta in prep note that this system hosts a known embedded protoplanet, as evidenced by kinematic signatures (Pinte et al., 2019). As discussed in Section 3.4, the interplay between an embedded planet and gravitational instability can suppress the spiral structure of the disc, hiding the signatures of the instability (Rowther et al., 2020).

## 4.2 The case of Elias 2–27

Elias 2–27 represents one of the most compelling targets for investigating gravitational instability (GI). Orbiting an M0-type star, its protoplanetary disc is one of the most massive observed to date. Initial estimates by Andrews et al. (2009), based on unresolved dust continuum emission, indicated a disc-to-star mass ratio of ~0.3. Subsequently, Pérez et al. (2016) provided the first resolved millimeter observations of the system using ALMA (left panel Figure 8), revealing two large-scale spiral arms and confirming the high dust mass previously reported. The presence of these prominent spiral features, coupled with the exceptionally high disc-to-star mass ratio, suggests that gravitational instability is a primary physical mechanism operating within the system.

TABLE 2 Systems with available dynamical star and disc mass measurements. In the table we report the best fit value from rotation curve modelling of stellar mass, disc mass, disc scale radius, the resulting disc-to-star mass ratio, the minimum Toomre paramter and the gas-to-dust ratio.

| Source | $M_\star$ [$M_\odot$] | $M_d$ [$M_\odot$] | $R_c$ [au] | $M_d/M_\star$ | min($Q$) | g/d | Ref. | Notes |
|---|---|---|---|---|---|---|---|---|
| Elias 2–27 | 0.46 | 0.08 | 200 | 0.17 | 4.5 | 80 | Veronesi et al. (2021), Longarini et al. (2024) | Scale radius is fixed at 200au, vertically isothermal model |
| AS 209 | 1.311 | $2\times10^{-5}$ | 126 | | | < 192 | Martire et al. (2024) | Disc mass compatible with zero |
| GM Aur | 1.128 | 0.118 | 96 | 0.10 | 4.9 | 240 | Martire et al. (2024) | |
| HD 163296 | 1.948 | 0.134 | 91 | 0.07 | 17.5 | 202 | Martire et al. (2024), Pezzotta et al. (2025) | Multi-molecular fit (7 molecules) |
| IM Lup | 1.194 | 0.106 | 115 | 0.09 | 5.5 | 134 | Martire et al. (2024) | |
| MWC 480 | 2.027 | 0.150 | 128 | 0.07 | 18.2 | 108 | Martire et al. (2024) | |
| AA tau | $0.624^{+0.033}_{-0.035}$ | $0.155^{+0.036}_{-0.036}$ | $156^{+106}_{-41}$ | $0.25^{+0.06}_{-0.06}$ | 2.6 | $1417^{+326}_{-330}$ | Longarini et al. (2025a) | Unreliable disc mass estimate |
| DM tau | $0.468^{+0.014}_{-0.015}$ | $0.057^{+0.019}_{-0.020}$ | $240^{+42}_{-27}$ | $0.12^{+0.04}_{-0.04}$ | 7.4 | $367^{+131}_{-121}$ | Longarini et al. (2025a) | |
| HD 34282 | $1.520^{+0.025}_{-0.031}$ | $0.143^{+0.045}_{-0.041}$ | $370^{+109}_{-78}$ | $0.09^{+0.03}_{-0.03}$ | 7.5 | $137^{+40}_{-43}$ | Longarini et al. (2025a) | |
| J1615 | $1.105^{+0.011}_{-0.012}$ | $0.082^{+0.014}_{-0.014}$ | $167^{+20}_{-15}$ | $0.07^{+0.01}_{-0.01}$ | 7.8 | $279^{+49}_{-49}$ | Longarini et al. (2025a) | |
| J1842 | $1.042^{+0.010}_{-0.011}$ | $0.078^{+0.013}_{-0.014}$ | $231^{+102}_{-50}$ | $0.07^{+0.01}_{-0.01}$ | 8.6 | $759^{+134}_{-129}$ | Longarini et al. (2025a) | |
| J1852 | $1.022^{+0.021}_{-0.021}$ | $0.044^{+0.024}_{-0.032}$ | $87^{+69}_{-16}$ | $0.04^{+0.02}_{-0.03}$ | 11.4 | $420^{+300}_{-301}$ | Longarini et al. (2025a) | |
| LkCa15 | $1.118^{+0.013}_{-0.015}$ | $0.108^{+0.016}_{-0.016}$ | $150^{+12}_{-16}$ | $0.10^{+0.01}_{-0.01}$ | 6.6 | $339^{+49}_{-50}$ | Longarini et al. (2025a) | |
| PDS 66 | $1.299^{+0.036}_{-0.101}$ | $0.038^{+0.099}_{-0.035}$ | $28^{+12}_{-5}$ | $0.03^{+0.08}_{-0.03}$ | 8.5 | $364^{+342}_{-357}$ | Longarini et al. (2025a) | Disc mass compatible with zero |
| SY Cha | $0.812^{+0.037}_{-0.041}$ | $0.084^{+0.044}_{-0.043}$ | $112^{+21}_{-15}$ | $0.10^{+0.05}_{-0.05}$ | 5.7 | $517^{+266}_{-273}$ | Longarini et al. (2025a) | |
| V4046 Sgr | $1.777^{+0.005}_{-0.006}$ | $0.058^{+0.006}_{-0.006}$ | $99^{+5}_{-5}$ | $0.03^{+0.00}_{-0.00}$ | 12.2 | $540^{+60}_{-59}$ | Longarini et al. (2025a) | Spectroscopic binary |
| HD 97048 | $2.226^{+0.054}_{-0.049}$ | $0.300^{+0.055}_{-0.061}$ | $172^{+24}_{-14}$ | $0.13^{+0.03}_{-0.03}$ | 4.7 | 640 | Pezzotta et al. (2026) | |
| WaOph6 | $0.956^{+0.006}_{-0.006}$ | $0.210^{+0.045}_{-0.038}$ | $647^{+193}_{-155}$ | $0.22^{+0.05}_{-0.04}$ | 5.4 | 1500 | Pezzotta et al. (2026) | |

References: 1 Veronesi et al. (2021); 2 Longarini et al. (2024); 3 Martire et al. (2024); 4 Pezzotta et al. (2025); 5 Longarini et al. (2025a); 6 Pezzotta et al. (2025).

Meru et al. (2017) investigated the origin of the spiral structure through 3D SPH simulations, evaluating three distinct scenarios: an embedded Jovian planet (internal companion), a stellar-mass companion at a large separation (external companion), and gravitational instability (GI). Their results demonstrated that an internal companion alone cannot replicate the observed morphology, effectively ruling it out as the sole mechanism in Elias 2–27. Conversely, both the GI and external companion models successfully reproduced the spiral features. Given the high disc mass, the authors suggest that the disc might be gravitationally unstable or may have undergone fragmentation, with the resulting fragment acting as the external companion driving the spirals.

To further discriminate between these scenarios, Forgan et al. (2018b) mapped the positions of the spiral density maxima to evaluate the radial dependence of the pitch angle. While an external companion is expected to produce asymmetric spirals with a pitch angle that deviates from a constant value, the observed morphology of Elias 2–27 exhibits a high degree of symmetry. These symmetric features lend further support to the GI hypothesis. Furthermore, Hall et al. (2018) used synthetic ALMA imaging to demonstrate that GI-generated spirals are only observable within a narrow parameter space of low opacity and efficient cooling. They also noted that significant dust trapping is required to produce the observed emission contrast, implying that the visibility of these structures depends heavily on the coupling between gas and dust.

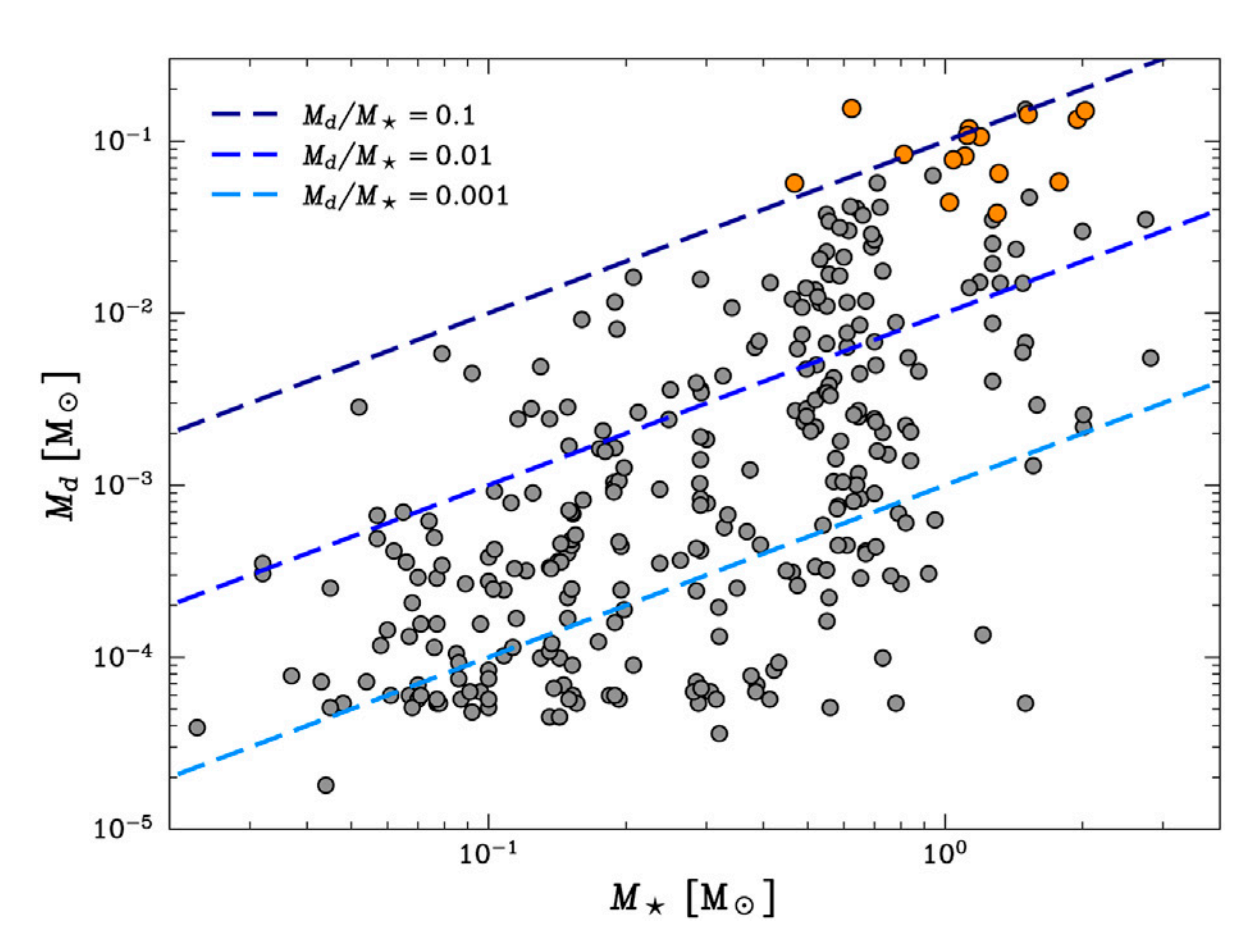

FIGURE 7
Disc mass as a function of stellar mass for the protoplanetary disc population. Grey dots represent the Manara et al. (2023) PPVII sample, where disc masses are derived from ALMA Band 6 or 7 continuum fluxes assuming a constant gas-to-dust ratio of 100. Orange dots denote the dynamical disc mass estimates compiled in Table 2. The blue dashed lines corresponds to a disc-to-star mass ratio of 0.1, 0.01, 0.001. This plot clearly shows the limitations of this method: it only works for massive discs, probing a very biased population.

Recent ALMA observations from the DSHARP large program (Huang et al., 2018b) have provided unprecedented high-resolution images of the dust continuum in Elias 2–27. This work characterised the spiral morphology with exceptional clarity and confirmed the presence of a gap at ~70$au$, suggesting that a planetary companion may coexist with the mechanism driving the spiral arms. Similar symmetric spiral structures were also identified in two other DSHARP discs, IM Lup and WaOph 6. Kinematic analysis of molecular line emission by Pinte et al. (2020) revealed enhanced non-Keplerian signatures co-located with the Elias 2–27 gap; however, the system's complex dynamics and the limited spectral resolution of the DSHARP data precluded definitive conclusions.

To address these limitations, Paneque-Carreño et al. (2021) presented high-sensitivity and high-spectral-resolution gas observations (right panel Figure 8). They identified global kinematic perturbations co-located with the dust spirals, which reminded of "GI wiggles" (Hall et al., 2020), providing further evidence for gravitational instability. Furthermore, Paneque-Carreño et al. (2021) reported a tentative detection of grain growth within the spiral arms, consistent with theoretical models of particle trapping (Dipierro et al., 2015). These kinematic data were later used by Veronesi et al. (2021) to perform the first rotation curve analysis aimed at constraining the dynamical mass of a protoplanetary disc. Their results yield a disc-to-star mass ratio of ~0.17 and a Toomre parameter of $Q \sim 4$, which, while above the classical threshold, is considered sufficiently low to support GI as a plausible origin for the observed features.

Building upon these results, Longarini et al. (2024) applied the analytical framework for "GI wiggles" developed in Longarini et al. (2021) to constrain angular momentum transport in Elias 2–27. By adopting the dynamical disc mass determined by Veronesi et al. (2021), the authors were able to break the degeneracy between disc mass and viscosity, finding that an effective $\alpha$–viscosity of $\alpha = 0.038$ accurately reproduces the amplitude of the observed kinematic perturbations. Furthermore, by assuming the self-similar surface density profile from Veronesi et al. (2021) and the temperature profile from Pérez et al. (2016), they calculated the predicted mass accretion rate onto the central star. The inferred value form the kinematic modelling is $\log_{10}\dot{M}_\star = -6.9 \pm 0.16$, and the observed value is $\log_{10}\dot{M}_\star = -7.2 \pm 0.5$. The remarkably good agreement between their model and the observed accretion rate suggests that the angular momentum transport driven by GI-induced spirals is sufficient to account for the stellar accretion in this system. Moreover, Longarini et al. (2024) point out that the disc-to-star mass ratio and the effective $\alpha$-viscosity are within the regime where dust particles can undergo direct gravitational collapse inside the gas spiral arms, provided the Stokes number is sufficiently large. The predicted mass of these dust fragments is on the order of $1-10\mathrm{M}_\oplus$, a range that is remarkably consistent with the mass estimate for a potential planetary companion inferred from the width of the observed dust gap.

Elias 2–27 is a significant example of how different signatures of gravitational instability line up in a self consistent and coherent way.

## 4.3 The case of AB Aur

AB Aurigae (AB Aur) is one of the most studied and debated systems in the context of gravitational instability. It is a Herbig Ae star of intermediate mass ($M_\star = 2.4\,M_\odot$) with an estimated age of 2.5–4.4 Myr (Garufi et al., 2024). Despite its relatively late evolutionary stage as a Class II young stellar object, its disc is exceptionally rich in gas and shows a complex morphology, making it one of the most compelling targets for studying the connection between disc self-gravity and planet formation.

The disc around AB Aur has been studied extensively across a wide range of wavelengths. Early near-infrared observations revealed prominent large-scale spiral arms in scattered light (Fukagawa et al., 2004), subsequently confirmed and characterised with ALMA (Boccaletti et al., 2020). The system also harbours a large central dust cavity as well as dust asymmetry (Tang et al., 2017). The origin of these substructures has been the subject of considerable debate, with proposed explanations including planetary companions, central binary, and gravitational instability. Currie et al. (2022) used the Subaru Telescope and the Hubble Space Telescope to find evidence for a Jovian protoplanet, AB Aur b, orbiting at a wide projected separation of ~93 au, possibly responsible for multiple planet-induced features in the disc. Cadman et al. (2021) assessed the formation history of this candidate planet, finding that core accretion timescales for such a massive object at wide separation typically exceed the system age, while gravitational instability, which operates on timescales of $10^4$–$10^5$ yr, remains a viable formation channel. Using SPH simulations, they estimated a critical disc mass for fragmentation of $M_{\mathrm{d},crit} = 0.3\,M_\odot$.

The most direct kinematic evidence for gravitational instability in AB Aur was presented by Speedie et al. (2024), as shown in Figure 9. Using deep ALMA observations of $^{13}$CO and $C^{18}O$ line emission, they identified kinematic signatures strongly resembling the predictions of GI simulations and analytical models (Hall et al., 2020; Longarini et al., 2021). From quantitative comparisons, they inferred a disc mass of up to one third of the stellar mass enclosed within the observed region, placing AB Aur firmly in the regime where self-gravity cannot be neglected. The observed kinematic perturbations are consistent with the large-scale GI wiggle (Longarini et al., 2021), and are distinct from the localised signatures expected from planet-disc interactions alone. Independent constraints on the disc mass come from Stapper et al. (2024), who used CO isotopologue observations of a large sample of Herbig discs to estimate their gas masses via thermochemical modelling, finding the AB Aur disc mass to lie at the upper boundary of the model grid, consistent with a self-gravitating regime.

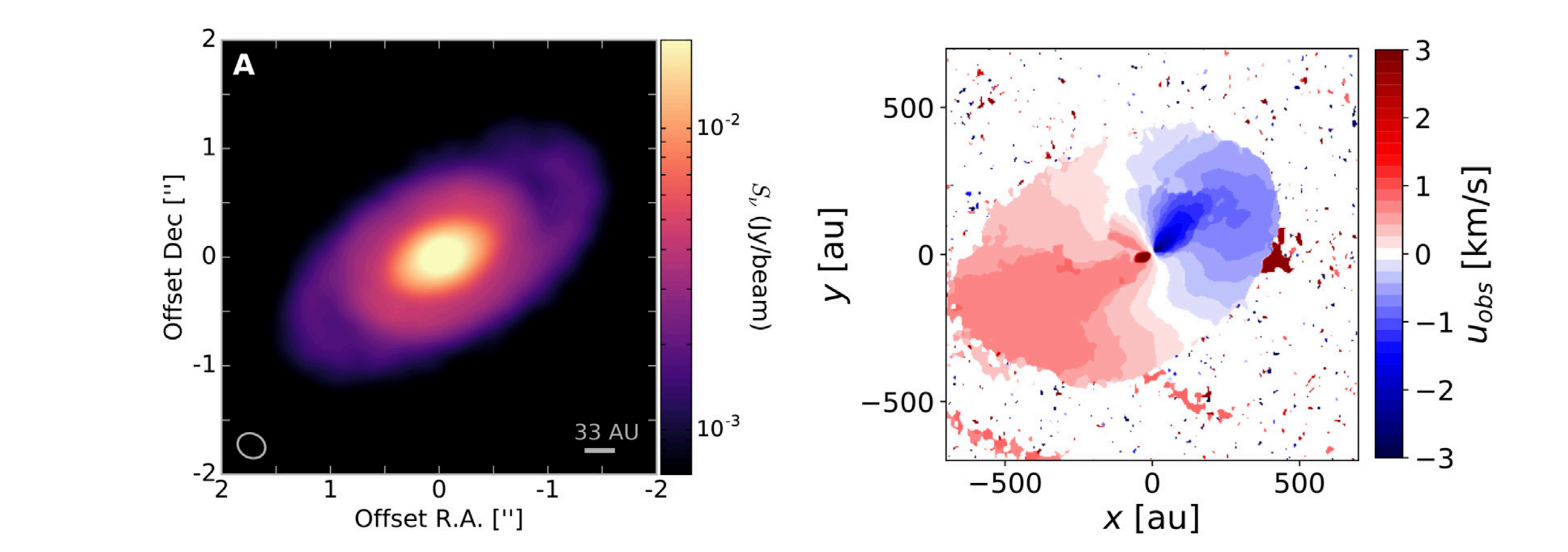

FIGURE 8
Dust and gas emission in the Elias 2–27 system. Left panel, **(a)**: Resolved millimetre continuum emission from ALMA (Pérez et al., 2016), showing the dust disc component. The image clearly reveals the two prominent large-scale symmetric spiral arms and the inner gap. Right panel: Map of $^{13}$CO molecular line emission (Paneque-Carreño et al., 2021), highlighting global non-Keplerian kinematics features. These perturbations align closely with the position of the dust spiral arms, and the morphology of the emission is reminiscent of "GI wiggles".

However, AB Aur is not an isolated system. Speedie et al. (2025) used deep ALMA $^{12}$CO and SO observations to map the large-scale kinematics. By separating disc-like from non-disc-like motions, they identified three large-scale streamers extending to ~1600 au in radius, interpreted as late infall from the surrounding environment. Where these infalling structures merge with the disc, a brightness enhancement in SO emission, a known shock tracer, is detected at $2.5\times$ the azimuthal average, providing direct evidence of shocks at the streamer-disc interface. This picture, in which late infall replenishes the disc, triggers or sustains gravitational instability, and modifies the conditions for planet formation, connects AB Aur directly to the infall-driven GI framework discussed in Section 3.5.

## 4.4 The case of IM Lup

IM Lup is a young pre-main sequence star (~1 Myr) located in the Lupus star-forming region. It hosts an unusually large disc, extending out to ≈300 au in the dust continuum and out to ≈1000 au in the gas (Cleeves et al., 2016), with a dynamical stellar mass of approximately a solar mass. The dust continuum emission shows clear evidence of a spiral morphology, which may be triggered by gravitational instability (Huang et al., 2018b). Cleeves et al. (2016) first estimated the disc mass from millimetre visibilities, finding a massive disc of $\sim 0.2\,M_\odot$. Verrios et al. (2022) proposed instead that the spiral structure could be generated by an embedded protoplanet. Through SPH simulations of planet-disc interactions, post-processed to reproduce CO, dust, and scattered light emission, they found that a high disc mass of $\sim 0.1\,M_\odot$ is nonetheless required to match the scattered light image, so that sub-micron sized grains remain well coupled in the upper disc layers.

A more direct constraint on the disc mass was obtained by Lodato et al. (2023) and Martire et al. (2024), who applied rotation curve modelling to the molecular line kinematics of IM Lup. Their analysis yields a disc-to-star mass ratio of ~0.1, and a minimum Toomre parameter of $Q \sim 4$, close enough to unity to place IM Lup in a regime where gravitational instability is plausible, even if not guaranteed. Unlike Elias 2–27 and AB Aur, the kinematics of IM Lup do not show evident non-Keplerian perturbations, making it difficult to conclusively distinguish between gravitational instability and planet-driven spirals as the origin of the observed substructure.

This degeneracy could in principle be broken by studying the time evolution of the spiral pattern. Spiral arms induced by a companion are long-lived, coherent structures that rotate at the orbital angular velocity of the companion, maintaining a constant pattern speed over time. In contrast, spirals generated by gravitational instability are transient: their instantaneous pattern speed is set by the local Keplerian velocity of the disc material, meaning that, if maintained, they would rapidly wind up due to differential rotation (Cossins et al., 2009). Instead, GI spirals continuously form, evolve, and dissipate on dynamical timescales. Monitoring the temporal evolution of the spiral morphology in IM Lup, for example, through multi-epoch ALMA observations, could therefore provide a powerful discriminant between the two scenarios. Yoshida et al. (2025) identified a winding motion of the spiral arms in dust continuum emission (see Figure 10), that is consistent with the Keplerian rotation of the disc, in line with the predictions of gravitationally unstable discs.

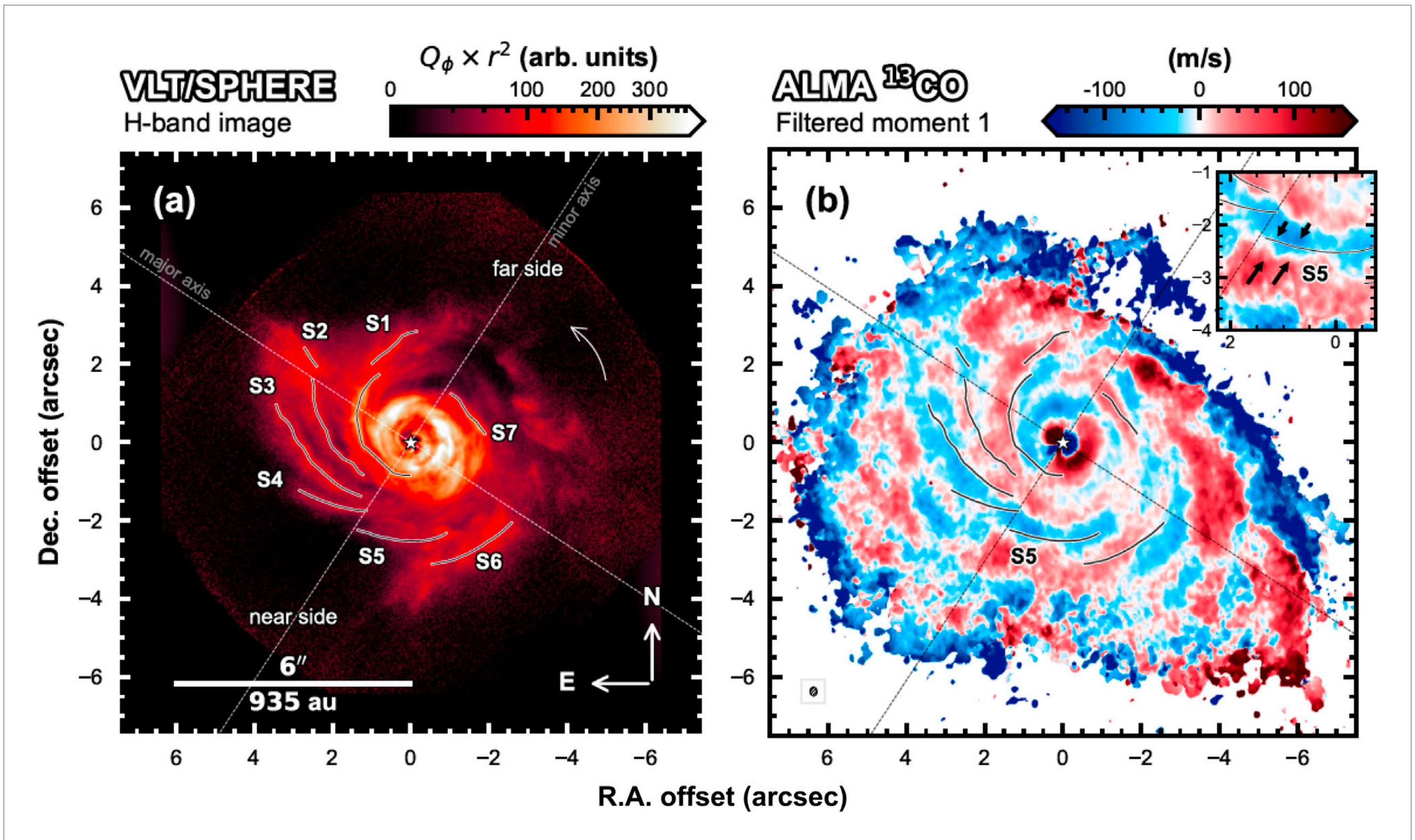

FIGURE 9
Left panel, **(a)**: VLT/SPHERE H-band scattered light image of the AB Aur disc (Boccaletti et al., 2020) tracing spiral structure in (sub-)micron-sized dust grains. Right panel, **(b)**: Filtered ALMA $^{13}$CO intensity-weighted mean velocity (moment 1) map, revealing residual gas motion within the bulk flow (Speedie et al., 2024).

## 5 Conclusion and future perspectives

Over the past decade, the study of gravitational instability in protoplanetary discs has undergone a profound transformation. The advent of ALMA has provided, for the first time, the angular resolution and sensitivity required to spatially resolve the kinematic and morphological signatures of self-gravity in individual discs. Simultaneously, advances in numerical algorithms and computational resources have enabled hydrodynamical simulations of unprecedented level of detail, incorporating increasingly sophisticated treatments of thermodynamics, dust dynamics, and the coupling between large-scale infall and disc evolution. Together, these developments have shifted the field from a largely theoretical exercise, where gravitational instability was invoked primarily as a formation mechanism for giant planets and brown dwarfs, to an observationally grounded research area where the physical conditions for GI can be directly measured, tested, and compared against theoretical predictions with quantitative precision.

On the theoretical side, significant progress has been made across several fronts. The long-standing debate over the critical cooling timescale for fragmentation has been resolved (Deng et al., 2017), recovering the original value of $\beta_{crit} = 3$ proposed by Gammie (2001) and establishing a firm upper limit on the angular momentum transport efficiency of gravito-turbulence. The role of thermodynamics has been greatly refined, moving beyond the idealised $\beta$-cooling prescription toward simulations incorporating local radiative transfer approximations (Young et al., 2024), full radiation transport (Xu et al., 2025a), and live Monte Carlo radiative transfer (Rowther et al., 2024). These studies have clarified the central role of stellar irradiation in regulating, and in some cases suppressing, the self-regulated state of GI (Cadman et al., 2020; Leedham et al., 2025). Building on the pioneering work of Rice et al. (2004), (2006), the connection between gravitational instability and the solid component of the disc has been substantially advanced: GI-induced spiral arms act as efficient dust traps, concentrating solid material and triggering the direct gravitational collapse of the dust layer into planetary-mass cores (Longarini et al., 2023a; Baehr et al., 2022; Rice et al., 2025), establishing gravitational instability as a viable mechanism for the rapid formation of planetary cores in the earliest stages of disc evolution. Finally, the picture of GI as a process driven solely by internal cooling has given way to a more complete framework in which infall from the parental envelope plays a central role (Kuffmeier et al., 2018), sustaining gravitational instability over extended timescales and establishing a tight coupling between disc mass, stellar mass, and accretion rate (Longarini et al., 2025b).

On the observational side, it is worth noting that gravitationally unstable discs do not represent the majority of systems observed by ALMA. This is partly a selection effect: the bulk of high-resolution ALMA surveys (Andrews et al., 2018) have preferentially targeted relatively evolved, Class II discs, with typical ages exceeding 1 Myr, at which point the disc mass has already decreased substantially through accretion onto the central star and the role of gravitational instability is expected to be marginal (Tobin et al., 2020). The youngest, most massive discs, Class 0 and Class I sources, where GI should be at play, remain comparatively under-sampled at high angular resolution, and even when observed are often subject to severe dust optical depth effects, which can hide spiral substructure even in discs that are young and massive enough to be gravitationally unstable (Ohashi et al., 2023). At the same time, we acknowledge that this observational bias cannot fully account for the absence of spiral structure in some of the youngest, most deeply embedded and massive systems that have been imaged at sufficient resolution. In these cases, the lack of observed spirals may instead reflect genuine thermodynamic stabilisation of the disc.

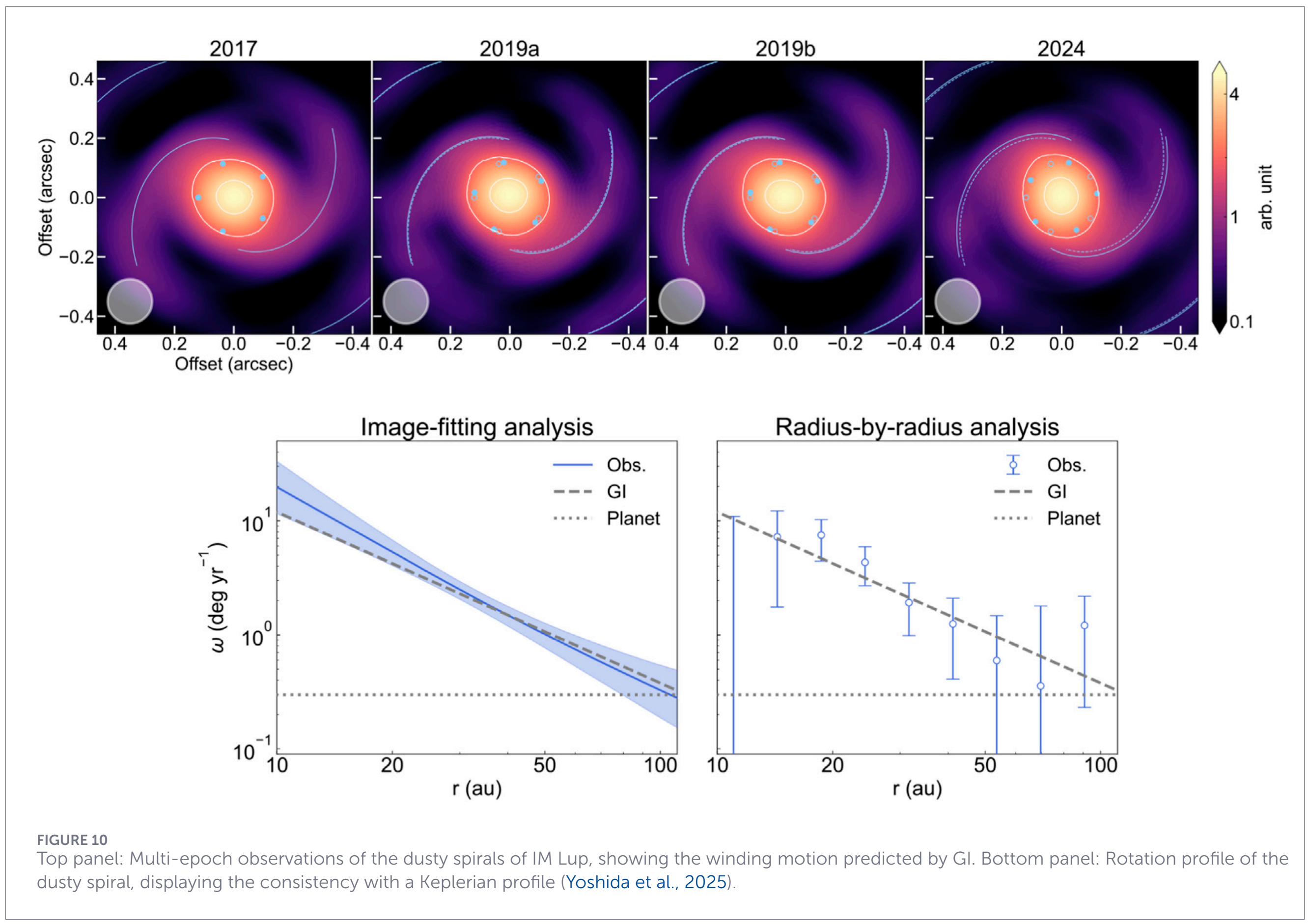

FIGURE 10
Top panel: Multi-epoch observations of the dusty spirals of IM Lup, showing the winding motion predicted by GI. Bottom panel: Rotation profile of the dusty spiral, displaying the consistency with a Keplerian profile (Yoshida et al., 2025).

Nevertheless, within this sample, the dynamical mass measurement technique has allowed us to identify a population of exceptionally massive discs whose kinematics deviate measurably from pure Keplerian rotation (Veronesi et al., 2021; Martire et al., 2024; Longarini et al., 2025a). For the most massive systems, spatially resolved kinematic perturbations consistent with GI-induced spiral structure have been detected, providing observational evidence that gravitational instability is an active physical process in these discs. Three systems, Elias 2–27, AB Aur, and IM Lup, stand out as particularly compelling cases, where multiple independent lines of evidence, including spiral morphology, non-Keplerian kinematics, and high disc-to-star mass ratios, converge toward a coherent GI picture.

## 5.1 Future perspectives

Despite the significant progress reviewed here, several fundamental questions remain open. In the following, we outline the most pressing theoretical and observational challenges that we expect will drive the field forward in the coming years.

### 5.1.1 Theoretical perspective

- A significant theoretical gap concerns the regime of very high disc-to-star mass ratios, $M_d/M_\star \gtrsim 0.5$, expected during the earliest phases of disc assembly when infall from the parental envelope is most vigorous. The majority of theoretical work on gravitational instability to date has focused on relatively light discs, where the pseudo-viscous approximation holds and angular momentum transport can be described as a local process. In this regime, the effective $\alpha$ viscosity provides a convenient and well-tested framework for characterising the self-regulated state of GI. However, as first shown by Lodato and Rice (2004), this local approximation breaks down when the disc-to-star mass ratio approaches unity. In this extreme regime, the gravitational potential of the disc becomes comparable to that of the central star, low-order global spiral modes dominate the dynamics, and angular momentum transport becomes inherently non-local. The physics of this regime draws direct parallels with the dynamics of self-gravitating galactic discs, where global modes and non-local transport have been extensively studied (Bertin, 2000), and where the theoretical tools developed in that context may prove valuable. Despite its astrophysical importance, this high-mass regime remains poorly understood in the context of protoplanetary discs. Early studies by Lodato and Rice (2005) have shown that for massive discs, the behaviour becomes more dynamical and is characterized by a series of recurrent phases of spiral activity. However, these studies were restricted to the simplest $\beta$ cooling prescription and much remains to be understood in this context. Understanding what dynamical and observational imprints this early, globally unstable phase leaves on the subsequent disc structure and on the conditions for planet formation represents an open questions in the field.

- A second major theoretical frontier concerns infall-driven gravitational instability. While the cooling-driven regime has been studied extensively over the past 2 decades, the infall-driven counterpart has only recently begun to receive the attention it deserves (Longarini et al., 2025b; Schib et al., 2025; Speedie et al., 2024), despite being the dominant mechanism during the Class 0 and I phases. Several fundamental questions remain open. From a thermodynamic perspective, the self-regulation mechanism in infall-driven discs has so far been characterised only in the isothermal limit (Longarini et al., 2025b); understanding how a more realistic treatment of disc thermodynamics modifies the self-regulated state is a critical next step. A second open question concerns the angular momentum budget of the infalling material itself. The outcome of mass loading depends sensitively on how and where material is injected into the disc: whether it arrives in a coherent streamer or distributed over an extended region, whether it is aligned or misaligned with the disc angular momentum vector. These factors determine not only the strength of the induced instability, but also the morphology of the resulting spiral structure and whether warps or misalignments are excited. Finally, a full understanding of infall-driven GI requires coupling high-resolution disc models with large-scale simulations of molecular cloud collapse (Kuffmeier et al., 2018; Lebreuilly et al., 2024), so that the properties of the infalling flow are self-consistently determined by the environment rather than prescribed as boundary conditions. Such coupled models represent the most promising path toward a complete picture of how the large-scale environment shapes the conditions for planet formation in the earliest stages of disc evolution.
- A third open question at the intersection of infall and planet formation regards the dust dynamics and solid core formation in infall-driven discs. In the cooling-driven regime, GI spiral arms are efficient dust traps precisely because their pattern speed is close to corotation (Cossins et al., 2009), minimising the relative velocity between the spiral pattern and the dust particles and allowing efficient concentration of solid material. However, in infall-driven discs the situation is different. Longarini et al. (2025b) showed that the pattern speed of infall-driven spirals adjusts to the angular momentum of the accreting material. This departure from corotation increases the relative velocity between the spiral arms and the dust grains, potentially suppressing the trapping efficiency and making the conditions for dust-driven gravitational collapse significantly harder to satisfy. Understanding whether and under what conditions dust concentration and core formation remain viable in infall-driven discs is therefore a critical open question, with direct implications for the timescale and efficiency of planet formation during the earliest phase of disc evolution.

### 5.1.2 Observational perspective

- On the observational side, a primary frontier is the study of younger, more embedded sources, where gravitational instability is expected to be most prevalent. The systems discussed in this paper, Elias 2–27, AB Aur, and IM Lup, are relatively evolved Class II objects; the conditions for GI are most naturally met in the earlier Class 0 and I phases, when the disc mass is highest and infall from the envelope is still ongoing. However, observing these systems with ALMA is challenging. The surrounding envelope makes them optically thick at millimetre wavelengths, obscuring the disc structure and complicating the interpretation of both continuum and line emission. The eDisk collaboration (Ohashi et al., 2023) has recently made significant progress in this direction, providing high-resolution images of young embedded discs, but also highlighting the difficulties in disentangling disc substructure from envelope contamination and optical depth effects. Beyond morphology, a deeper characterisation of the physical properties of young embedded sources, including their accretion rates, disc-to-star mass ratios, and thermal structures, is essential to test whether the conditions for infall-driven GI are met.
- A second observational priority is the enlargement of the dynamical mass sample. The method has so far been successfully applied to 18 systems, but the sample remains limited by strict observational requirements: the disc must be spatially extended, bright in molecular line emission, and sufficiently massive to produce a measurable deviation from Keplerian rotation. These constraints naturally favour large, gas-rich discs around intermediate-mass stars, leaving compact discs and young embedded sources largely unexplored. Extending the method to these regimes would be particularly valuable, as compact discs may harbour significant self-gravity despite their small apparent size, and embedded sources are precisely those where GI is expected to be most active. This will likely require combining the rotation curve approach with complementary mass tracers to break the degeneracies that arise when the self-gravitating signal is weak or the thermal structure is poorly constrained.
- A third observational priority concerns the role of gravitational instability from a population perspective. A complete understanding of GI requires moving beyond individual case studies toward a statistical characterisation of its imprints across the disc population. Key questions include: how common are self-gravitating discs at different evolutionary stages? What fraction of the observed substructure in young discs, rings, gaps, spirals, and asymmetries, can be attributed to gravitational instability rather than planet-disc interactions? And, most fundamentally, what are the initial conditions that GI establishes for subsequent planet formation, in terms of disc mass distribution, dust grain growth, and the spatial distribution of solid material? Answering these questions requires combining large, homogeneous disc surveys with population synthesis models that self-consistently incorporate the role of self-gravity, infall, and dust evolution from the earliest phases of disc assembly (Schib et al., 2025). Such an approach would allow us to connect the detailed physics of GI studied in individual systems to the broader statistics of planetary system architectures emerging from exoplanet surveys.

## Author contributions

CL: Conceptualization, Writing – original draft, Writing – review and editing. GL: Conceptualization, Writing – original draft.

## Funding

The author(s) declared that financial support was received for this work and/or its publication. CL has been supported by the UK Science and Technology Research Council (STFC) via the consolidated grant ST/W000997/1. GL acknowledges financial support from the project “Space Observations of Proto-planetary Accretion Disc Evolution (SPADE)” in the framework of the “PROgramma di RIcerca Spaziale di base (PRORIS)”, financed by the Italian Ministry of University and Research (Ministero dell’Università e della Ricerca; MUR). The authors thank the University of Cambridge for providing the financial support to publish this paper.

## Acknowledgments

The authors thank Cathie Clarke, James Rogers, Alice Somigliana for useful discussions and comments.

## Conflict of interest

The author(s) declared that this work was conducted in the absence of any commercial or financial relationships that could be construed as a potential conflict of interest.

The handling editor ŁT declared a past co-authorship with the author GL.

## Generative AI statement

The author(s) declared that generative AI was not used in the creation of this manuscript.

Any alternative text (alt text) provided alongside figures in this article has been generated by Frontiers with the support of artificial intelligence and reasonable efforts have been made to ensure accuracy, including review by the authors wherever possible. If you identify any issues, please contact us.

## Publisher’s note

## References

ALMA Partnership, Brogan, C. L., Pérez, L. M., Hunter, T. R., Dent, W. R. F., Hales, A. S., et al. (2015). The 2014 ALMA long baseline campaign: first results from high angular resolution observations toward the HL Tau Region. *ApJL* 808, L3. doi: 10.1088/2041-8205/808/1/L3

Andrews, S. M. (2020). Observations of protoplanetary disk structures. *ARAA* 58, 483–528. doi: 10.1146/annurev-astro-031220-010302

Andrews, S. M., and Williams, J. P. (2007). A submillimeter view of circumstellar dust disks in $\rho$ ophiuchi. *ApJ* 671, 1800–1812. doi: 10.1086/522885

Andrews, S. M., Wilner, D. J., Hughes, A. M., Qi, C., and Dullemond, C. P. (2009). Protoplanetary disk structures in ophiuchus. *ApJ* 700, 1502–1523. doi: 10.1088/0004-637X/700/2/1502

Andrews, S. M., Huang, J., Pérez, L. M., Isella, A., Dullemond, C. P., Kurtovic, N. T., et al. (2018). The disk substructures at high angular resolution project (DSHARP). I. Motivation, sample, calibration, and overview. *ApJL* 869, L41. doi: 10.3847/2041-8213/aaf741

Andrews, S. M., Teague, R., Wirth, C. P., Huang, J., and Zhu, Z. (2024). On kinematic measurements of self-gravity in protoplanetary disks. *ApJ* 970, 153. doi: 10.3847/1538-4357/ad5285

Baehr, H., Zhu, Z., and Yang, C.-C. (2022). Direct Formation of planetary embryos in self-gravitating disks. *ApJ* 933, 100. doi: 10.3847/1538-4357/ac7228

Baehr, H., Rice, K., Yang, C.-C., and Hall, C. (2025). On the gravitational collapse of small dust grains in self-gravitating disk structures. *ApJ* 995, 39. doi: 10.3847/1538-4357/ae1a73

Baruteau, C., Meru, F., and Paardekooper, S.-J. (2011). Rapid inward migration of planets formed by gravitational instability. *MNRAS* 416, 1971–1982. doi: 10.1111/j.1365-2966.2011.19172.x

Bate, M. R. (2018). On the diversity and statistical properties of protostellar discs. *MNRAS* 475, 5618–5658. doi: 10.1093/mnras/sty169

Benitez-Llambay, P. (2026). Planet disk interaction and evolution. *Encycl. Astrophysics* 1, 250–269. doi: 10.1016/B978-0-443-21439-4.00033-X

Bergin, E. A., Cleeves, L. I., Gorti, U., Zhang, K., Blake, G. A., Green, J. D., et al. (2013). An old disk still capable of forming a planetary system. *Nature* 493, 644–646. doi: 10.1038/nature11805

Bertin, G. (2000). Dynamics of galaxies

Bertin, G., and Lodato, G. (1999). A class of self-gravitating accretion disks. *A&A* 350, 694–704. doi: 10.48550/arXiv.astro-ph/9908095

Bertin, G., and Romeo, A. B. (1988). Global spiral modes in stellar disks containing gas. *A&A* 195, 105–113.

Birnstiel, T. (2024). Dust growth and evolution in protoplanetary disks. *ARAA* 62, 157–202. doi: 10.1146/annurev-astro-071221-052705

Blandford, R. D., and Payne, D. G. (1982). Hydromagnetic flows from accretion disks and the production of radio jets. *MNRAS* 199, 883–903. doi: 10.1093/mnras/199.4.883

Boccaletti, A., Di Folco, E., Pantin, E., Dutrey, A., Guilloteau, S., Tang, Y. W., et al. (2020). Possible evidence of ongoing planet formation in AB aurigae. A showcase of the SPHERE/ALMA synergy. *A&A* 637, L5. doi: 10.1051/0004-6361/202038008

Bollati, F., Lodato, G., Price, D. J., and Pinte, C. (2021). The theory of kinks - I. A semi-analytic model of velocity perturbations due to planet-disc interaction. *MNRAS* 504, 5444–5454. doi: 10.1093/mnras/stab1145

Booth, R. A., and Clarke, C. J. (2016). Collision velocity of dust grains in self-gravitating protoplanetary discs. *MNRAS* 458, 2676–2693. doi: 10.1093/mnras/stw488

Booth, R. A., and Clarke, C. J. (2019). Characterizing gravito-turbulence in 3D: turbulent properties and stability against fragmentation. *MNRAS* 483, 3718–3729. doi: 10.1093/mnras/sty3340

Boss, A. P. (1997). Giant planet formation by gravitational instability. *Science* 276, 1836–1839. doi: 10.1126/science.276.5320.1836

Cadman, J., Rice, K., Hall, C., Haworth, T. J., and Biller, B. (2020). Fragmentation favoured in discs around higher mass stars. *MNRAS* 492, 5041–5051. doi: 10.1093/mnras/staa187

Cadman, J., Rice, K., and Hall, C. (2021). AB Aurigae: possible evidence of planet formation through the gravitational instability. *MNRAS* 504, 2877–2888. doi: 10.1093/mnras/stab905

Casassus, S., van der Plas, G. M., Perez, S., Dent, W. R. F., Fomalont, E., Hagelberg, J., et al. (2013). Flows of gas through a protoplanetary gap. *Nature* 493, 191–194. doi: 10.1038/nature11769

Clarke, C. J. (2009). Pseudo-viscous modelling of self-gravitating discs and the formation of low mass ratio binaries. *MNRAS* 396, 1066–1074. doi: 10.1111/j.1365-2966.2009.14774.x

Clarke, C. J., Tazzari, M., Juhasz, A., Rosotti, G., Booth, R., Facchini, S., et al. (2018). High-resolution Millimeter Imaging of the CI Tau Protoplanetary Disk: a Massive Ensemble of Protoplanets from 0.1 to 100 au. *ApJL* 866, L6. doi: 10.3847/2041-8213/aae36b

Cleeves, L. I., Öberg, K. I., Wilner, D. J., Huang, J., Loomis, R. A., Andrews, S. M., et al. (2016). The coupled physical structure of gas and dust in the IM Lup protoplanetary disk. *ApJ* 832, 110. doi: 10.3847/0004-637X/832/2/110

Cossins, P., Lodato, G., and Clarke, C. J. (2009). Characterizing the gravitational instability in cooling accretion discs. *MNRAS* 393, 1157–1173. doi: 10.1111/j.1365-2966.2008.14275.x

Curone, P., Facchini, S., Andrews, S. M., Testi, L., Benisty, M., Czekala, I., et al. (2025). exoALMA. IV. Substructures, asymmetries, and the faint outer disk in continuum emission. *ApJL* 984, L9. doi: 10.3847/2041-8213/adc438

Currie, T., Lawson, K., Schneider, G., Lyra, W., Wisniewski, J., Grady, C., et al. (2022). Images of embedded Jovian planet formation at a wide separation around AB Aurigae. *Nat. Astron.* 6, 751–759. doi: 10.1038/s41550-022-01634-x

Deng, H., Mayer, L., and Meru, F. (2017). Convergence of the critical cooling rate for protoplanetary disk fragmentation achieved: the key role of numerical dissipation of angular momentum. *ApJ* 847, 43. doi: 10.3847/1538-4357/aa872b

Dipierro, G., Pinilla, P., Lodato, G., and Testi, L. (2015). Dust trapping by spiral arms in gravitationally unstable protostellar discs. *MNRAS* 451, 974–986. doi: 10.1093/mnras/stv970

Dong, R., Zhu, Z., Rafikov, R. R., and Stone, J. M. (2015a). Observational signatures of planets in protoplanetary disks: spiral arms observed in scattered light imaging can be induced by planets. *ApJL* 809, L5. doi: 10.1088/2041-8205/809/1/L5

Dong, R., Zhu, Z., and Whitney, B. (2015b). Observational signatures of planets in protoplanetary disks I. Gaps opened by single and multiple young planets in disks. *ApJ* 809, 93. doi: 10.1088/0004-637X/809/1/93

Drkazkowska, J., Bitsch, B., Lambrechts, M., Mulders, G. D., Harsono, D., Vazan, A., et al. (2023). Planet formation theory in the era of ALMA and kepler: from pebbles to exoplanets. In *Protostars and Planets VII*, eds. S. Inutsuka, Y. Aikawa, T. Muto, K. Tomida, and M. Tamura doi: 10.48550/arXiv.2203.09759

Durisen, R. H., Boss, A. P., Mayer, L., Nelson, A. F., Quinn, T., and Rice, W. K. M. (2007). "Gravitational instabilities in gaseous protoplanetary disks and implications for giant planet formation," in *Protostars and Planets V*. Editors B. Reipurth, D. Jewitt, and K. Keil, 607. doi: 10.48550/arXiv.astro-ph/0603179

Forgan, D., and Rice, K. (2011). The Jeans mass as a fundamental measure of self-gravitating disc fragmentation and initial fragment mass. *MNRAS* 417, 1928–1937. doi: 10.1111/j.1365-2966.2011.19380.x

Forgan, D., and Rice, K. (2013). Towards a population synthesis model of objects formed by self-gravitating disc fragmentation and tidal downsizing. *MNRAS* 432, 3168–3185. doi: 10.1093/mnras/stt672

Forgan, D. H., Hall, C., Meru, F., and Rice, W. K. M. (2018a). Towards a population synthesis model of self-gravitating disc fragmentation and tidal downsizing II: the effect of fragment-fragment interactions. *MNRAS* 474, 5036–5048. doi: 10.1093/mnras/stx2870

Forgan, D. H., Ilee, J. D., and Meru, F. (2018b). Are Elias 2-27's spiral arms driven by self-gravity, or by a companion? A comparative spiral morphology study. *ApJL* 860, L5. doi: 10.3847/2041-8213/aac7c9

Fukagawa, M., Hayashi, M., Tamura, M., Itoh, Y., Hayashi, S. S., Oasa, Y., et al. (2004). Spiral structure in the circumstellar disk around AB aurigae. *ApJL* 605, L53–L56. doi: 10.1086/420699

Galloway-Sprietsma, M., Bae, J., Izquierdo, A. F., Stadler, J., Longarini, C., Teague, R., et al. (2025). exoALMA. V. Gaseous emission surfaces and temperature structures. *ApJL* 984, L10. doi: 10.3847/2041-8213/adc437

Gammie, C. F. (2001). Nonlinear outcome of gravitational instability in cooling, gaseous disks. *ApJ* 553, 174–183. doi: 10.1086/320631

Garufi, A., Ginski, C., van Holstein, R. G., Benisty, M., Manara, C. F., Pérez, S., et al. (2024). The SPHERE view of the Taurus star-forming region. The full census of planet-forming disks with GTO and DESTINYS programs. *A&A* 685, A53. doi: 10.1051/0004-6361/202347586

Gundlach, B., and Blum, J. (2015). The stickiness of micrometer-sized water-ice particles. *ApJ* 798, 34. doi: 10.1088/0004-637X/798/1/34

Guzmán, V. V., Huang, J., Andrews, S. M., Isella, A., Pérez, L. M., Carpenter, J. M., et al. (2018). The disk substructures at High Angular Resolution Program (DSHARP). VIII. The rich ringed substructures in the AS 209 disk. *ApJL* 869, L48. doi: 10.3847/2041-8213/aaedae

Hall, C., Rice, K., Dipierro, G., Forgan, D., Harries, T., and Alexander, R. (2018). Is the spiral morphology of the Elias 2-27 circumstellar disc due to gravitational instability? *MNRAS* 477, 1004–1014. doi: 10.1093/mnras/sty550

Hall, C., Dong, R., Teague, R., Terry, J., Pinte, C., Paneque-Carreño, T., et al. (2020). Predicting the kinematic evidence of gravitational instability. *ApJ* 904, 148. doi: 10.3847/1538-4357/abac17

Haworth, T. J., Cadman, J., Meru, F., Hall, C., Albertini, E., Forgan, D., et al. (2020). Massive discs around low-mass stars. *MNRAS* 494, 4130–4148. doi: 10.1093/mnras/staa883

Huang, J., Andrews, S. M., Dullemond, C. P., Isella, A., Pérez, L. M., Guzmán, V. V., et al. (2018a). The disk substructures at high angular resolution Project (DSHARP). II. Characteristics of annular substructures. *ApJL* 869, L42. doi: 10.3847/2041-8213/aaf740

Huang, J., Andrews, S. M., Pérez, L. M., Zhu, Z., Dullemond, C. P., Isella, A., et al. (2018b). The disk substructures at high angular resolution Project (DSHARP). *ApJL* 869, L43. doi: 10.3847/2041-8213/aaf7a0

Ilee, J. D., Forgan, D. H., Evans, M. G., Hall, C., Booth, R., Clarke, C. J., et al. (2017). The chemistry of protoplanetary fragments formed *via* gravitational instabilities. *MNRAS* 472, 189–204. doi: 10.1093/mnras/stx1966

Isella, A., Pérez, L. M., Carpenter, J. M., Ricci, L., Andrews, S., and Rosenfeld, K. (2013). An Azimuthal Asymmetry in the LkH$\alpha$ 330 disk. *ApJ* 775, 30. doi: 10.1088/0004-637X/775/1/30

Izquierdo, A. F., Stadler, J., Galloway-Sprietsma, M., Benisty, M., Pinte, C., Bae, J., et al. (2025). exoALMA. III. Line-intensity modeling and system property extraction from protoplanetary disks. *ApJL* 984, L8. doi: 10.3847/2041-8213/adc439

Jog, C. J., and Solomon, P. M. (1984). Two-fluid gravitational instabilities in a galactic disk. *ApJ* 276, 114–126. doi: 10.1086/161597

Kratter, K., and Lodato, G. (2016). Gravitational instabilities in circumstellar disks. *ARAA* 54, 271–311. doi: 10.1146/annurev-astro-081915-023307

Kratter, K. M., and Matzner, C. D. (2006). Fragmentation of massive protostellar discs. *MNRAS* 373, 1563–1576. doi: 10.1111/j.1365-2966.2006.11103.x

Kratter, K. M., Matzner, C. D., and Krumholz, M. R. (2008). Global models for the evolution of embedded, accreting protostellar disks. *ApJ* 681, 375–390. doi: 10.1086/587543

Kratter, K. M., Matzner, C. D., Krumholz, M. R., and Klein, R. I. (2010). On the role of disks in the formation of stellar systems: a numerical parameter study of rapid accretion. *ApJ* 708, 1585–1597. doi: 10.1088/0004-637X/708/2/1585

Kuffmeier, M., Frimann, S., Jensen, S. S., and Haugbølle, T. (2018). Episodic accretion: the interplay of infall and disc instabilities. *MNRAS* 475, 2642–2658. doi: 10.1093/mnras/sty024

Kuffmeier, M., Dullemond, C. P., Reissl, S., and Goicovic, F. G. (2021). Misaligned disks induced by infall. *A&A* 656, A161. doi: 10.1051/0004-6361/202039614

Lebreuilly, U., Hennebelle, P., Colman, T., Maury, A., Tung, N. D., Testi, L., et al. (2024). Synthetic populations of protoplanetary disks: impact of magnetic fields and radiative transfer. *A&A* 682, A30. doi: 10.1051/0004-6361/202346558

Leedham, C. S., Booth, R. A., and Clarke, C. J. (2025). Effect of irradiation model on 2D hydrodynamic simulations of self-gravitating protoplanetary discs. *MNRAS* 539, 2780–2789. doi: 10.1093/mnras/staf644

Lin, M.-K., and Kratter, K. M. (2016). On the gravitational stability of Gravito-turbulent accretion disks. *ApJ* 824, 91. doi: 10.3847/0004-637X/824/2/91

Lin, C. C., and Shu, F. H. (1964). On the spiral structure of disk galaxies. *ApJ* 140, 646. doi: 10.1086/147955

Lodato, G., and Clarke, C. J. (2011). Resolution requirements for smoothed particle hydrodynamics simulations of self-gravitating accretion discs. *MNRAS* 413, 2735–2740. doi: 10.1111/j.1365-2966.2011.18344.x

Lodato, G., and Rice, W. K. M. (2004). Testing the locality of transport in self-gravitating accretion discs. *MNRAS* 351, 630–642. doi: 10.1111/j.1365-2966.2004.07811.x

Lodato, G., and Rice, W. K. M. (2005). Testing the locality of transport in self-gravitating accretion discs - II. The massive disc case. *MNRAS* 358, 1489–1500. doi: 10.1111/j.1365-2966.2005.08875.x

Lodato, G., Dipierro, G., Ragusa, E., Long, F., Herczeg, G. J., Pascucci, I., et al. (2019). The newborn planet population emerging from ring-like structures in discs. *MNRAS* 486, 453–461. doi: 10.1093/mnras/stz913

Lodato, G., Rampinelli, L., Viscardi, E., Longarini, C., Izquierdo, A., Paneque-Carreño, T., et al. (2023). Dynamical mass measurements of two protoplanetary discs. *MNRAS* 518, 4481–4493. doi: 10.1093/mnras/stac3223

Löhnert, L., Krätschmer, S., and Peeters, A. G. (2020). Saturation mechanism and generated viscosity in gravito-turbulent accretion disks. *A&A* 640, A53. doi: 10.1051/0004-6361/202038023

Lombardi, J. C., McInally, W. G., and Faber, J. A. (2015). An efficient radiative cooling approximation for use in hydrodynamic simulations. *MNRAS* 447, 25–35. doi: 10.1093/mnras/stu2432

Long, F., Pinilla, P., Herczeg, G. J., Harsono, D., Dipierro, G., Pascucci, I., et al. (2018). Gaps and Rings in an ALMA Survey of disks in the taurus star-forming Region. *ApJ* 869, 17. doi: 10.3847/1538-4357/aae8e1

Longarini, C., Lodato, G., Toci, C., Veronesi, B., Hall, C., Dong, R., et al. (2021). Investigating protoplanetary disk cooling through kinematics: analytical GI wiggle. *ApJL* 920, L41. doi: 10.3847/2041-8213/ac2df6

Longarini, C., Armitage, P. J., Lodato, G., Price, D. J., and Ceppi, S. (2023a). The role of the drag force in the gravitational stability of dusty planet-forming disc - II. Numerical simulations. *MNRAS* 522, 6217–6235. doi: 10.1093/mnras/stad1400

Longarini, C., Lodato, G., Bertin, G., and Armitage, P. J. (2023b). The role of the drag force in the gravitational stability of dusty planet forming disc - I. Analytical theory. *MNRAS* 519, 2017–2029. doi: 10.1093/mnras/stac3653

Longarini, C., Lodato, G., Clarke, C. J., Speedie, J., Paneque-Carreño, T., Arrigoni, E., et al. (2024). Angular momentum transport *via* gravitational instability in the Elias 2-27 disc. *A&A* 686, L6. doi: 10.1051/0004-6361/202450187

Longarini, C., Lodato, G., Rosotti, G., Andrews, S., Winter, A., Stadler, J., et al. (2025a). exoALMA. XII. Weighing and sizing exoALMA disks with rotation curve modelling. *ApJL* 984, L17. doi: 10.3847/2041-8213/adc431

Longarini, C., Price, D. J., Kratter, K. M., Lodato, G., and Clarke, C. J. (2025b). Infall-driven gravitational instability in accretion discs. *MNRAS* 541, 1145–1163. doi: 10.1093/mnras/staf1018

Malik, M., Meru, F., Mayer, L., and Meyer, M. (2015). On the gap-opening criterion of migrating planets in protoplanetary disks. *ApJ* 802, 56. doi: 10.1088/0004-637X/802/1/56

Manara, C. F., Ansdell, M., Rosotti, G. P., Hughes, A. M., Armitage, P. J., Lodato, G., et al. (2023). Demographics of young stars and their protoplanetary disks: lessons learned on disk evolution and its connection to planet Formation. In *Protostars and Planets VII*, eds. S. Inutsuka, Y. Aikawa, T. Muto, K. Tomida, and M. Tamura doi: 10.48550/arXiv.2203.09930

Marino, S., Casassus, S., Perez, S., Lyra, W., Roman, P. E., Avenhaus, H., et al. (2015). Compact dust concentration in the MWC 758 protoplanetary disk. *ApJ* 813, 76. doi: 10.1088/0004-637X/813/1/76

Marois, C., Macintosh, B., Barman, T., Zuckerman, B., Song, I., Patience, J., et al. (2008). Direct imaging of multiple planets orbiting the star HR 8799. *Science* 322, 1348–1352. doi: 10.1126/science.1166585

Martire, P., Longarini, C., Lodato, G., Rosotti, G. P., Winter, A., Facchini, S., et al. (2024). Rotation curves in protoplanetary disks with thermal stratification. Physical model and observational evidence in MAPS disks. *A&A* 686, A9. doi: 10.1051/0004-6361/202348546

Mayor, M., and Queloz, D. (1995). A Jupiter-mass companion to a solar-type star. *Nature* 378, 355–359. doi: 10.1038/378355a0

Meru, F., and Bate, M. R. (2010). Exploring the conditions required to form giant planets *via* gravitational instability in massive protoplanetary discs. *MNRAS* 406, 2279–2288. doi: 10.1111/j.1365-2966.2010.16867.x

Meru, F., and Bate, M. R. (2011). Non-convergence of the critical cooling time-scale for fragmentation of self-gravitating discs. *MNRAS* 411, L1–L5. doi: 10.1111/j.1745-3933.2010.00978.x

Meru, F., and Bate, M. R. (2012). On the convergence of the critical cooling time-scale for the fragmentation of self-gravitating discs. *MNRAS* 427, 2022–2046. doi: 10.1111/j.1365-2966.2012.22035.x

Meru, F., Juhász, A., Ilee, J. D., Clarke, C. J., Rosotti, G. P., and Booth, R. A. (2017). On the origin of the spiral morphology in the Elias 2-27 circumstellar disk. *ApJL* 839, L24. doi: 10.3847/2041-8213/aa6837

Miotello, A., van Dishoeck, E. F., Williams, J. P., Ansdell, M., Guidi, G., Hogerheijde, M., et al. (2017). Lupus disks with faint CO isotopologues: low gas/dust or high carbon depletion? *A&A* 599, A113. doi: 10.1051/0004-6361/201629556

Miotello, A., Kamp, I., Birnstiel, T., Cleeves, L. C., and Kataoka, A. (2023). Setting the stage for planet Formation: measurements and implications of the fundamental disk properties. In *Protostars and Planets VII*, eds. S. Inutsuka, Y. Aikawa, T. Muto, K. Tomida, and M. Tamura doi: 10.48550/arXiv.2203.09818

Musiolik, G. (2021). Growth of aggregates with liquid-like ice shells in protoplanetary discs. *MNRAS* 506, 5153–5159. doi: 10.1093/mnras/stab1963

Musiolik, G., and Wurm, G. (2019). Contacts of water ice in protoplanetary Disks—laboratory experiments. *ApJ* 873, 58. doi: 10.3847/1538-4357/ab0428

Nayakshin, S. (2010). Formation of planets by tidal downsizing of giant planet embryos. *MNRAS* 408, L36–L40. doi: 10.1111/j.1745-3933.2010.00923.x

Nayakshin, S., Zhang, L., Ćalović, A., Lee, H., Baruteau, C., Meru, F., et al. (2026). Disc fragmentation - II. Ejection of low-mass free-floating planets from growing binary systems. *MNRAS* 546, stag043. doi: 10.1093/mnras/stag043

Ohashi, N., Tobin, J. J., Jørgensen, J. K., Takakuwa, S., Sheehan, P., Aikawa, Y., et al. (2023). Early planet Formation in embedded disks (eDisk). I. Overview of the program and first results. *ApJ* 951, 8. doi: 10.3847/1538-4357/acd384

Paardekooper, S., Dong, R., Duffell, P., Fung, J., Masset, F. S., Ogilvie, G., et al. (2023). *Planet-Disk Interactions and Orbital Evolution. in Protostars and Planets VII*, eds. S. Inutsuka, Y. Aikawa, T. Muto, K. Tomida, and M. Tamura doi: 10.48550/arXiv.2203.09595

Paneque-Carreño, T., Pérez, L. M., Benisty, M., Hall, C., Veronesi, B., Lodato, G., et al. (2021). Spiral arms and a massive dust disk with Non-Keplerian kinematics: possible evidence for gravitational instability in the disk of Elias 2-27. *ApJ* 914, 88. doi: 10.3847/1538-4357/abf243

Pérez, L. M., Carpenter, J. M., Andrews, S. M., Ricci, L., Isella, A., Linz, H., et al. (2016). Spiral density waves in a young protoplanetary disk. *Science* 353, 1519–1521. doi: 10.1126/science.aaf8296

Pezzotta, V., Facchini, S., Longarini, C., Lodato, G., and Martire, P. (2025). The two-dimensional pressure structure of the HD 163296 protoplanetary disk as probed by multi-molecule kinematics. *A&A* 694, A108. doi: 10.1051/0004-6361/202451307

Pezzotta, V., Facchini, S., Izquierdo, A. F., Lodato, G., Longarini, C., Bae, J., et al. (2026). Extending dynamical mass measurements: probing GI as a possible origin of mm-dust spirals. *arXiv e-prints*. doi: 10.48550/arXiv.2607.15923

Pinte, C., van der Plas, G., Ménard, F., Price, D. J., Christiaens, V., Hill, T., et al. (2019). Kinematic detection of a planet carving a gap in a protoplanetary disk. *Nat. Astron.* 3, 1109–1114. doi: 10.1038/s41550-019-0852-6

Pinte, C., Price, D. J., Ménard, F., Duchêne, G., Christiaens, V., Andrews, S. M., et al. (2020). Nine localized deviations from Keplerian rotation in the DSHARP circumstellar disks: kinematic evidence for protoplanets carving the gaps. *ApJL* 890, L9. doi: 10.3847/2041-8213/ab6dda

Pringle, J. E. (1981). Accretion discs in astrophysics. *ARAA* 19, 137–162. doi: 10.1146/annurev.aa.19.090181.001033

Rafikov, R. R. (2005). Can giant planets form by direct gravitational instability? *ApJL* 621, L69–L72. doi: 10.1086/428899

Rice, W. K. M., Armitage, P. J., Bate, M. R., and Bonnell, I. A. (2003). The effect of cooling on the global stability of self-gravitating protoplanetary discs. *MNRAS* 339, 1025–1030. doi: 10.1046/j.1365-8711.2003.06253.x

Rice, W. K. M., Lodato, G., Pringle, J. E., Armitage, P. J., and Bonnell, I. A. (2004). Accelerated planetesimal growth in self-gravitating protoplanetary discs. *MNRAS* 355, 543–552. doi: 10.1111/j.1365-2966.2004.08339.x

Rice, W. K. M., Lodato, G., and Armitage, P. J. (2005). Investigating fragmentation conditions in self-gravitating accretion discs. *MNRAS* 364, L56–L60. doi: 10.1111/j.1745-3933.2005.00105.x

Rice, W. K. M., Lodato, G., Pringle, J. E., Armitage, P. J., and Bonnell, I. A. (2006). Planetesimal formation *via* fragmentation in self-gravitating protoplanetary discs. *MNRAS* 372, L9–L13. doi: 10.1111/j.1745-3933.2006.00215.x

Rice, W. K. M., Armitage, P. J., Mamatsashvili, G. R., Lodato, G., and Clarke, C. J. (2011). Stability of self-gravitating discs under irradiation. *MNRAS* 418, 1356–1362. doi: 10.1111/j.1365-2966.2011.19586.x

Rice, K., Baehr, H., Young, A. K., Booth, R., Rowther, S., Meru, F., et al. (2025). Dust density enhancements and the direct formation of planetary cores in gravitationally unstable discs. *MNRAS* 539, 3421–3435. doi: 10.1093/mnras/staf714

Riols, A., Roux, B., Latter, H., and Lesur, G. (2020). Dust dynamics and vertical settling in gravitoturbulent protoplanetary discs. *MNRAS* 493, 4631–4642. doi: 10.1093/mnras/staa567

Rowther, S., and Meru, F. (2020). Planet migration in self-gravitating discs: survival of planets. *MNRAS* 496, 1598–1609. doi: 10.1093/mnras/staa1590

Rowther, S., Meru, F., Kennedy, G. M., Nealon, R., and Pinte, C. (2020). Hiding signatures of gravitational instability in protoplanetary disks with planets. *ApJL* 904, L18. doi: 10.3847/2041-8213/abc704

Rowther, S., Nealon, R., and Meru, F. (2022). Warping away gravitational instabilities in protoplanetary discs. *ApJ* 925, 163. doi: 10.3847/1538-4357/ac3975

Rowther, S., Nealon, R., and Meru, F. (2023). Continuing to hide signatures of gravitational instability in protoplanetary discs with planets. *MNRAS* 518, 763–773. doi: 10.1093/mnras/stac3106

Rowther, S., Nealon, R., Meru, F., Wurster, J., Aly, H., Alexander, R., et al. (2024). The role of drag and gravity on dust concentration in a gravitationally unstable disc. *MNRAS* 528, 2490–2500. doi: 10.1093/mnras/stae167

Safronov, V. S. (1960). On the gravitational instability in flattened systems with axial symmetry and non-uniform rotation. *Ann. d'Astrophysique* 23, 979.

Schib, O., Mordasini, C., Emsenhuber, A., and Helled, R. (2025). DIPSY: a new disc instability population SYnthesis: II. The populations of companions formed through disc instability. *A&A* 704, A28. doi: 10.1051/0004-6361/202556261

Segura-Cox, D. M., Schmiedeke, A., Pineda, J. E., Stephens, I. W., Fernández-López, M., Looney, L. W., et al. (2020). Four annular structures in a protostellar disk less than 500,000 years old. *Nature* 586, 228–231. doi: 10.1038/s41586-020-2779-6

Sheehan, P. D., and Eisner, J. A. (2018). Multiple gaps in the disk of the class I protostar GY 91. *ApJ* 857, 18. doi: 10.3847/1538-4357/aaae65

Speedie, J., Dong, R., Hall, C., Longarini, C., Veronesi, B., Paneque-Carreño, T., et al. (2024). Gravitational instability in a planet-forming disk. *Nature* 633, 58–62. doi: 10.1038/s41586-024-07877-0

Speedie, J., Dong, R., Teague, R., Segura-Cox, D., Pineda, J. E., Calcino, J., et al. (2025). Mapping the merging Zone of late infall in the AB Aur planet-forming system. *ApJL* 981, L30. doi: 10.3847/2041-8213/adb7d5

Stamatellos, D., Whitworth, A. P., Bisbas, T., and Goodwin, S. (2007). Radiative transfer and the energy equation in SPH simulations of star formation. *A&A* 475, 37–49. doi: 10.1051/0004-6361:20077373

Stapper, L. M., Hogerheijde, M. R., van Dishoeck, E. F., Lin, L., Ahmadi, A., Booth, A. S., et al. (2024). Constraining the gas mass of Herbig disks using CO isotopologues. *A&A* 682, A149. doi: 10.1051/0004-6361/202347271

Tang, Y.-W., Guilloteau, S., Dutrey, A., Muto, T., Shen, B.-T., Gu, P.-G., et al. (2017). Planet formation in AB aurigae: imaging of the inner gaseous spirals observed inside the dust cavity. *ApJ* 840, 32. doi: 10.3847/1538-4357/aa6af7

Terry, J. P., Hall, C., Longarini, C., Lodato, G., Toci, C., Veronesi, B., et al. (2022). Constraining protoplanetary disc mass using the GI wiggle. *MNRAS* 510, 1671–1679. doi: 10.1093/mnras/stab3513

Tobin, J. J., Sheehan, P. D., Megeath, S. T., Díaz-Rodríguez, A. K., Offner, S. S. R., Murillo, N. M., et al. (2020). The VLA/ALMA nascent disk and Multiplicity (VANDAM) survey of orion protostars. II. A statistical characterization of class 0 and class I protostellar disks. *ApJ* 890, 130. doi: 10.3847/1538-4357/ab6f64

Toomre, A. (1964). On the gravitational stability of a disk of stars. *ApJ* 139, 1217–1238. doi: 10.1086/147861

van Capelleveen, R. F., Ginski, C., Kenworthy, M. A., Byrne, J., Lawlor, C., McLachlan, D., et al. (2025). WIde separation planets in time (WISPIT): a gap-clearing planet in a multi-ringed disk around the young solar-type star WISPIT 2. *ApJL* 990, L8. doi: 10.3847/2041-8213/adf721

Venturini, J., Ronco, M. P., and Guilera, O. M. (2020). Setting the stage: planet Formation and volatile delivery. *Space Sci. Rev.* 216, 86. doi: 10.1007/s11214-020-00700-y

Veronesi, B., Paneque-Carreño, T., Lodato, G., Testi, L., Pérez, L. M., Bertin, G., et al. (2021). A dynamical measurement of the disk mass in Elias 227. *ApJL* 914, L27. doi: 10.3847/2041-8213/abfe6a

Veronesi, B., Longarini, C., Lodato, G., Laibe, G., Hall, C., Facchini, S., et al. (2024). Weighing protoplanetary discs with kinematics: physical model, method, and benchmark. *A&A* 688, A136. doi: 10.1051/0004-6361/202348237

Verrios, H. J., Price, D. J., Pinte, C., Hilder, T., and Calcino, J. (2022). Kinematic evidence for an embedded planet in the IM Lupi disk. *ApJL* 934, L11. doi: 10.3847/2041-8213/ac7f44

Vorobyov, E. I., and Basu, S. (2005a). The effect of non-isothermality on the gravitational collapse of spherical clouds and the evolution of protostellar accretion. *MNRAS* 363, 1361–1368. doi: 10.1111/j.1365-2966.2005.09528.x

Vorobyov, E. I., and Basu, S. (2005b). The origin of episodic accretion bursts in the early stages of star Formation. *ApJL* 633, L137–L140. doi: 10.1086/498303

Vorobyov, E. I., and Basu, S. (2007). Self-regulated gravitational accretion in protostellar discs. *MNRAS* 381, 1009–1017. doi: 10.1111/j.1365-2966.2007.12321.x

Wada, K., Tanaka, H., Okuzumi, S., Kobayashi, H., Suyama, T., Kimura, H., et al. (2013). Growth efficiency of dust aggregates through collisions with high mass ratios. *A&A* 559, A62. doi: 10.1051/0004-6361/201322259

Walmswell, J., Clarke, C., and Cossins, P. (2013). The evolution of planetesimal swarms in self-gravitating protoplanetary discs. *MNRAS* 431, 1903–1913. doi: 10.1093/mnras/stt314

Xu, W., Jiang, Y.-F., Kunz, M. W., and Stone, J. M. (2025a). Global simulations of gravitational instability in protostellar disks with full radiation transport. I. Stochastic fragmentation with optical-depth-dependent rate and universal fragment mass. *ApJ* 986, 91. doi: 10.3847/1538-4357/add14a

Xu, W., Jiang, Y.-F., Kunz, M. W., and Stone, J. M. (2025b). Global simulations of gravitational instability in protostellar disks with full radiation transport. II. Locality of gravitoturbulence, clumpy spirals, and implications for observable substructure. *ApJ* 986, 92. doi: 10.3847/1538-4357/add14b

Yoshida, T. C., Nomura, H., Doi, K., Barraza-Alfaro, M., Teague, R., Furuya, K., et al. (2025). Winding motion of spirals in a gravitationally unstable protoplanetary disk. *Nat. Astron.* 9, 1672–1679. doi: 10.1038/s41550-025-02639-y

Young, A. K., Celeste, M., Booth, R. A., Rice, K., Koval, A., Carter, E., et al. (2024). Introducing two improved methods for approximating radiative cooling in hydrodynamical simulations of accretion discs. *MNRAS* 531, 1746–1755. doi: 10.1093/mnras/stae1249

Zhu, Z., Hartmann, L., Nelson, R. P., and Gammie, C. F. (2012). Challenges in forming planets by gravitational instability: disk irradiation and clump migration, accretion, and tidal destruction. *ApJ* 746, 110. doi: 10.1088/0004-637X/746/1/110